\documentclass[twocolumn,amsmath,aps,superscriptaddress,longbibliography]{revtex4-2}
\usepackage{array}[=2016-10-06]
\usepackage{graphicx}
\usepackage{bm}
\usepackage{xcolor}
\usepackage{subfigure}
\usepackage{hyperref}
\usepackage{booktabs}
\usepackage{siunitx}
\usepackage[version=4]{mhchem}

\usepackage{amssymb}
\usepackage{algorithm}
\usepackage{algpseudocode}
\usepackage{placeins}
\usepackage{float}
\usepackage{enumitem}
\hypersetup{colorlinks=true, citecolor=blue,pdfborder={0 0 0}}
\makeatletter
\renewcommand{\section}{\@startsection{section}{1}{0mm}
  {-\baselineskip}{0.5\baselineskip}{\bf\center}}
\makeatother
\setcitestyle{super}

\newcommand{\todo}[1]{%
  \expandafter\ifx\expandafter\relax\detokenize{#1}\relax%
    \textcolor{red}{[TODO]}%
  \else%
    \textcolor{red}{[TODO: #1]}%
  \fi%
}

\begin{document}
\title{Correlation-Converged Virtual Orbitals for Accurate and Efficient Quantum Molecular Simulations}

\author{Qian Wang}
\thanks{These authors contributed equally to this work.}
\affiliation{Department of Mechanical Engineering, City University of Hong Kong, Kowloon, Hong Kong 999077, China}

\author{Calvin Ku}
\thanks{These authors contributed equally to this work.}
\affiliation{Hon Hai Research Institute, Taipei, Taiwan}
\affiliation{Department of Mechanical Engineering, City University of Hong Kong, Kowloon, Hong Kong 999077, China}

\author{Jyh-Pin Chou}
\affiliation{Graduate School of Advanced Technology, National Taiwan University, Taipei 10617, Taiwan}
\affiliation{Physics Division, National Center for Theoretical Sciences, National Taiwan University, Taipei 106319, Taiwan}

\author{Peng-Jen Chen}
\email{pjchen1015@gmail.com}
\affiliation{Department of Mechanical Engineering, City University of Hong Kong, Kowloon, Hong Kong 999077, China}

\author{Alice Hu}
\email{alicehu@cityu.edu.hk}
\affiliation{Department of Mechanical Engineering, City University of Hong Kong, Kowloon, Hong Kong 999077, China}
\affiliation{Department of Material Science and Engineering, City University of Hong Kong, Kowloon, Hong Kong 999077, China}

\author{Min-Hsiu Hsieh}
\email{min-hsiu.hsieh@foxconn.com}
\affiliation{Hon Hai Research Institute, Taipei, Taiwan}

\begin{abstract}
  Density functional theory with plane-wave basis sets is widely employed in computational materials science, including applications to isolated molecular systems. However, the inadequate description of electron correlation remains a fundamental limitation. Accurate correlation treatments based on many-body Hamiltonians require reliable representations of both occupied and virtual orbitals, yet virtual orbitals are often poorly described in conventional computational schemes, resulting in reduced accuracy. In this work, we introduce localized correlation-converged virtual orbitals (LCCVOs) as an efficient basis for constructing accurate many-body Hamiltonians in molecular systems. Using a substantially reduced number of orbitals, the LCCVO framework yields dissociation energies for singlet, doublet, and triplet molecules that are comparable to, and in many cases exceed, those obtained with high-level correlation-consistent basis sets such as cc-pVXZ (X = D, T, Q, 5). These results demonstrate the efficiency, scalability, and robustness of the LCCVO approach for high-accuracy quantum chemical calculations.
\end{abstract}

\maketitle

\section*{Introduction}
The quantum many-body problem lies at the heart of electronic structure theory and remains computationally challenging, particularly for systems with strong electronic correlations \cite{abrams1997simulation}.
Quantum computing has been proposed as a potential route to overcome these limitations, with numerous algorithms designed for future fault-tolerant hardware that utilize a variety of basis sets and quantizations schemes\cite{kuOptimizingQuantumChemistry2026,berryQubitizationArbitraryBasis2019,vonburgQuantumComputingEnhanced2021,leeEvenMoreEfficient2021,georgesQuantumSimulationsChemistry2025}.
However, current and near-term quantum implementations remains constrained by a limited number of quantum bits (qubits), which in turn restrict the size of available basis sets\cite{bauer2020quantum}.
Most many-body methods implemented on near-term devices rely on second-quantized Hamiltonians using the molecular orbital (MO) basis, where the accuracy is strictly bottlenecked by the steep hardware cost required to represent many-electron wavefunctions in a finite basis\cite{bauer2020quantum, preskill2018quantum}.
As a result, quantum chemistry calculations are typically limited to minimal Gaussian basis sets, which are computationally efficient but often lack quantitative reliability.
While higher-level basis sets provide systematic convergence, their rapidly growing orbital spaces render them impractical for near-term quantum devices \cite{dunning1989gaussian}.

Plane-wave (PW) basis sets offer systematic convergence and a structured Hamiltonian that is particularly attractive from an algorithmic perspective. However, many-body Hamiltonians constructed from PW-derived eigenstates, such as Kohn-Sham (KS) orbitals obtained from density functional theory (DFT), often yield poor results even for simple molecules due to highly delocalized virtual orbitals that weakly couple to occupied states, leading to inefficient recovery of correlation energy \cite{song2023periodic}. Previous efforts, such as the correlation-optimized virtual orbital (COVO) method, demonstrated improved performance for small, spin-unpolarized systems but suffer from limited scalability and poor applicability to larger or spin-polarized molecules \cite{song2023periodic, bylaska2021quantum}.

In this work, we introduce localized correlation-converged virtual orbitals (LCCVOs), a plane-wave-based framework that preserves orbital localization while systematically improving the description of correlation energy. By optimizing virtual orbitals with respect to their interactions with occupied states, LCCVOs form compact active spaces that significantly reduce the number of orbitals required for accurate many-body calculations, with advantages that become increasingly pronounced for larger molecular systems. In addition, extrapolation to the complete basis set (CBS) limit and correction of the finite size effect are adopted.

\begin{table*}
  \caption{
    The ground state energies of diatomic molecules using different basis-sets along with the number of orbitals $N^{\text{orb}}$ used.
    Equilibrium bond length ($R_e$) and dissociation energy ($D_0$) are displayed in \AA and eV, respectively.
    The error in $D_0$ is represented in percentages.
  }
  \sisetup{detect-weight}
  \begin{tabular}{
    p{2cm}
    S[table-column-width=12mm, table-format=3.0]
    S[table-column-width=15mm, table-format=1.3]
    S[table-column-width=12mm, table-format=1.2]
    S[table-column-width=22mm, table-format=-2.2]
    p{1mm}
    S[table-column-width=15mm, table-format=3.0]
    S[table-column-width=15mm, table-format=1.3]
    S[table-column-width=12mm, table-format=1.2]
    S[table-column-width=22mm, table-format=-2.2]
  }
    \toprule
    Method & {$N^{\text{orb}}$} & {$R_e$} & {$E_0$} & {error in $D_0$} & & {$N^{\text{orb}}$} & {$R_e$} & {$E_0$} & {error in $D_0$} \\
    \cmidrule(lr){1-5} \cmidrule(lr){7-10}
    & \multicolumn{4}{c}{{\bf \ce{H2}} (ZPE = \qty{0.270}{eV})} & & \multicolumn{4}{c}{{\bf \ce{N2}} (ZPE = \qty{0.146}{eV})} \\
    \cmidrule(lr){2-5} \cmidrule(lr){7-10}
    \bf LCCVO  & \bf 15 & \bf 0.735 & \bf 4.73 & \bf -0.70 &  & \bf 40 & \bf 1.100 & \bf 9.69 & \bf -2.20 \\
    STO-3G     & 2      & 0.735     & 5.31     & 11.79     &  & 10     & 1.100     & 5.59     & -43.59    \\
    6-31G      & 4      & 0.746     & 4.22     & -11.16    &  & 18     & 1.200     & 6.62     & -33.20    \\
    cc-pVDZ    & 10     & 0.760     & 4.49     & -5.47     &  & 28     & 1.110     & 8.19     & -17.36    \\
    cc-pVTZ    & 28     & 0.743     & 4.70     & -1.05     &  & 60     & 1.110     & 8.94     & -9.79     \\
    cc-pVQZ    & 60     & 0.742     & 4.73     & -0.42     &  & 110    & 1.110     & 9.26     & -6.56     \\
    cc-pV5Z    & 110    & 0.742     & 4.74     & -0.21     &  & 182    & 1.110     & 9.37     & -5.45     \\
    CBS-limit  & {--}   & {--}      & 4.75     & 0.00      &  & {--}   & {--}      & 9.43     & -4.84     \\
    Experiment & {--}   & 0.741     & 4.75\cite{herzberg1970dissociation}     & {--}      &  & {--}   & 1.110     & 9.91\cite{lu2023observation,bytautas2005correlation}     & {--}       \\
    \midrule
    & \multicolumn{4}{c}{{\bf \ce{O2}} (ZPE = \qty{0.098}{eV})} & & \multicolumn{4}{c}{{\bf \ce{\ce{CN}}} (ZPE = \qty{0.128}{eV})} \\
    \cmidrule(lr){2-5} \cmidrule(lr){7-10}
    \bf LCCVO  & \bf 50 & \bf 1.200 & \bf 4.99 & \bf -4.32 &  & \bf 40 & \bf 1.200 & \bf 7.42 & \bf-5.74 \\
    STO-3G     & 10     & 1.280     & 2.50     & -52.11    &  & 10     & 1.200     & 5.41     & -31.26   \\
    6-31G      & 18     & 1.270     & 2.64     & -49.43    &  & 18     & 1.200     & 5.54     & -29.61   \\
    cc-pVDZ    & 28     & 1.210     & 3.97     & -23.95    &  & 28     & 1.170     & 6.33     & -19.57   \\
    cc-pVTZ    & 60     & 1.190     & 4.47     & -14.37    &  & 60     & 1.160     & 6.93     & -11.94   \\
    cc-pVQZ    & 110    & 1.190     & 4.69     & -10.15    &  & 110    & 1.160     & 7.19     & -8.64    \\
    cc-pV5Z    & 182    & 1.190     & 4.77     & -8.62     &  & 182    & 1.160     & 7.28     & -7.50    \\
    CBS-limit  & {--}   & {--}      & 4.82     & -7.66     &  & {--}   & {--}      & 7.32     & -6.99    \\
    Experiment & {--}   & 1.210     & 5.22\cite{wang2024bond,bytautas2005correlation}     & {--}      &  & {--}   & 1.170     & 7.87\cite{{pradhan1994dissociation,huang1992heat}}     & {--}     \\
    \midrule
    & \multicolumn{4}{c}{{\bf \ce{C2}} (ZPE = \qty{0.115}{eV})} & & \multicolumn{4}{c}{} \\
    \cmidrule(lr){2-5}
    \bf LCCVO  & \bf 40 & \bf 1.250 & \bf 5.53 & \bf -13.85 &  & \multicolumn{4}{c}{} \\
    STO-3G     & 10     & 1.250     & 5.33     & -16.98     &  & \multicolumn{4}{c}{} \\
    6-31G      & 18     & 1.250     & 4.74     & -26.17     &  & \multicolumn{4}{c}{} \\
    cc-pVDZ    & 28     & 1.270     & 4.79     & -25.39     &  & \multicolumn{4}{c}{} \\
    cc-pVTZ    & 60     & 1.270     & 5.20     & -19.00     &  & \multicolumn{4}{c}{} \\
    cc-pVQZ    & 110    & 1.240     & 5.37     & -16.36     &  & \multicolumn{4}{c}{} \\
    cc-pV5Z    & 182    & 1.240     & 5.43     & -15.42     &  & \multicolumn{4}{c}{} \\
    CBS-limit  & {--}   & {--}      & 5.46     & -14.95     &  & \multicolumn{4}{c}{} \\
    Experiment & {--}   & 1.243     & 6.42\cite{pradhan1994dissociation,su2011bonding,bytautas2005correlation}     & {--}       &  & \multicolumn{4}{c}{} \\
    \bottomrule
  \end{tabular}
  \label{result}
\end{table*}

\section*{Results}

The PW-derived LCCVO method is applied to investigate several molecules, including singlet (closed-shell \ce{H2} and \ce{N2} molecules and open-shell \ce{C2} molecule), doublet (\ce{CN}), and triplet (\ce{O2}) molecules. For comparison with experimental dissociation energies $D_0$, measured zero-point energy (ZPE) of each molecule is subtracted from the calculated electronic dissociation energy $D_e$: $D_0 = D_e - \text{ZPE}$ \cite{martin1997electric, wang2024bond, lu2023observation, herzberg1970dissociation, pradhan1994dissociation, huang1992heat, su2011bonding, bytautas2005correlation}. As listed in Table \ref{result}, it is apparent that the equilibrium bond length is insensitive to the basis set. Nonetheless, the dissociation energy show strong dependence on the basis set. Although STO-3G and 6-31G basis sets require few orbitals for calculations, they yield unsatisfactory dissociation energies for the simplest molecule \ce{H2}, let alone for more complex molecules such as \ce{N2}. The use of the cc-pVXZ bases enhances the accuracy at the cost of increasing number of required orbitals, with acceptable results obtained using the cc-pVQZ or cc-pV5Z bases. However, the large number of required orbitals makes the calculations inefficient and almost prohibits the application of near-term quantum computing to solve the problem. On the contrary, our LCCVO framework gives accurate dissociation energies with a moderate number of required orbitals. It is worthwhile to mention that the dissociation energies obtained by LCCVO are more accurate than that from cc-pVQZ and cc-pV5Z, except for the case of \ce{H2} where the latter two yield slightly higher accuracy. The advantage of the LCCVO framework becomes pronounced for molecules containing more electrons. For instance, the error in dissociation energy of \ce{N2} is only $-2.20\%$, surpassing the cc-pV5Z basis and even the extrapolated-CBS limit derived from the cc-pVXZ basis-sets. The total energy versus atomic distance is shown in Supplementary Figures 1-3.

Moreover, LCCVO can also be used to study open-shell molecules, in contrast to the original COVO method which only considers closed-shell molecules. As Table~\ref{result} shows, the dissociation energies of triplet \ce{O2} and doublet \ce{CN} from LCCVO are in good agreement with the experimental values. LCCVO also performs better than both the large cc-pV5Z basis and the extrapolated-CBS limit, with fewer number of required orbitals. At the end, we also apply our method to the singlet \ce{C2} molecule. Although some people consider the singlet ground state of \ce{C2} to be closed-shell, it behaves differently from typical closed-shell molecules. Despite obtaining the best calculated dissociation energies compared with experimental values, the error of the LCCVO method still exceeds $10\%$ as shown in Table \ref{result}. The relatively poor dissociation energy results of this unstable \ce{C2} molecule is discussed in Supplementary Note 1.

\section*{Discussion}

In pioneering works that employ quantum computing algorithms to study molecular properties, simple basis sets (such as STO-3G and 6-31G) are commonly adopted to construct the many-body Hamiltonians, which can be solved using either real quantum hardware or quantum simulators. However, the resulting dissociation energies are often unsatisfactory, even for the \ce{H2} molecule. Although the error, which is approximately $18\%$, is known to originate from the use of a minimal basis set, employing higher-level basis sets to improve the accuracy is impractical due to the limited number of available qubits. This limitation motivates the development of a numerical scheme that can effectively interface with quantum computing algorithms while preserving quantitative accuracy. Since the commonly adopted basis sets are relatively inflexible, we instead focus on a PW-based approach.

For this purpose, the first step is to identify the origin of the inaccurate total energies obtained from many-body Hamiltonians constructed using PW-derived orbitals. While dissociation energies computed directly from DFT are reasonably accurate, the corresponding many-body calculations based on the KS orbitals yield poor results. This discrepancy stems from the fact that virtual orbitals produced by DFT calculations are not explicitly optimized for correlation effects due to the density-dependent nature of DFT \cite{cohen2008fractional}. Furthermore, those virtual orbitals often include artificial vacuum states for isolated systems that require an explicit vacuum regions~\cite{medvedev2017density}. Detailed analysis reveals that this issue is closely related to the delocalization of certain virtual orbitals. Since molecular virtual orbitals should be intrinsically localized, the presence of such delocalized orbitals represents an artifact inherent to the plane-wave basis, which must be eliminated to achieve high accuracy. At the same time, the proposed numerical scheme must remain computationally efficient.

In light of the above issues, we propose a quantum computational framework that incorporate the concept of COVOs with an improved correlation-optimization procedure, together with the extrapolation to the CBS limit and correction of the finite size effect, called LCCVO. This approach yields improved molecular orbitals and enhances the fidelity of the resulting many-body Hamiltonians, enabling accurate determination of dissociation energies for both spin-unpolarized and spin-polarized molecular systems. The calculated results show excellent agreement with experimental data, while requiring significantly fewer orbitals. These findings highlight the critical role of basis-set design in quantum computational chemistry. The LCCVO framework achieves accuracy comparable to that of classical high-level basis sets while demanding substantially fewer quantum resources. As a result, it provides a scalable route toward simulations of larger and more complex molecular systems on near-term quantum hardware. Its close agreement with experimental values further underscores its reliability for practical applications.

\begin{figure*}[t]
\includegraphics[width=0.9\linewidth]{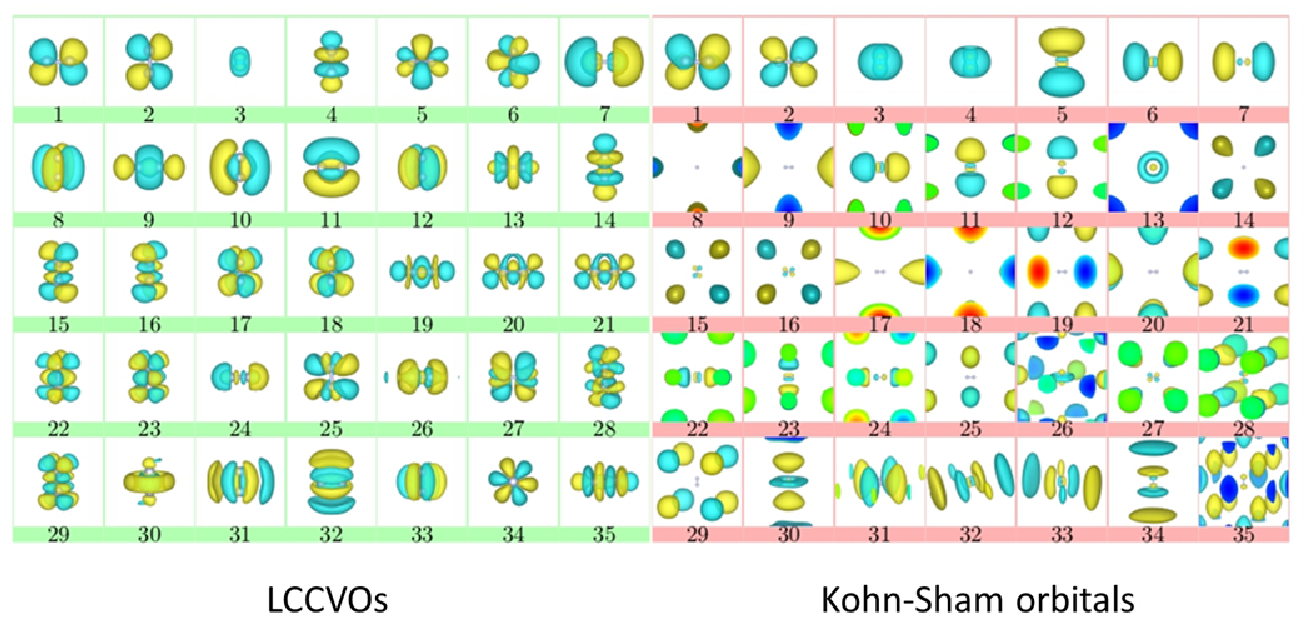}
\caption{
    Spatial distribution of LCCVOs and virtual Kohn-Sham orbitals of \ce{N2}.
    Comparison shows that the artificial vacuum states are effectively eliminated in the LCCVOs, resulting in virtual orbitals (orbitals 7-35) that are centred around the molecule.
}
\label{orbital}
\end{figure*}

Figure~\ref{orbital} compares the spatial distributions of 35 virtual orbitals of \ce{N2} obtained with and without our LCCVO framework. The virtual orbitals under LCCVO are clearly centred around the \ce{N2} molecule, whereas the virtual orbitals obtained directly from DFT are strongly mixed with the vacuum states, except for a few low-energy orbitals. Under these circumstances, correlation effects are expected to be poorly captured without LCCVO. The improved performance of the LCCVO may be attributed to their enhanced ability to extract correlation effects from orbitals that would otherwise be either weakly coupled or contaminated by vacuum character. These observations are consistent with previous benchmarks demonstrating the superiority of LCCVOs in systems characterized by strong correlation and complex orbital behaviours. Overall, the results highlight the methodological advantage of LCCVOs in accurately describing correlation energy contributions associated with virtual orbitals. The bar chart in Supplementary Figure 4 compares the correlation energies obtained with and without LCCVO for different numbers of virtual orbitals, where the LCCVO method recovers substantially more correlation energies, indicating a more effective treatment of electron correlation.

Furthermore, a closer examination of the results summarized in Table \ref{result} shows that most dissociation energies are slightly underestimated, indicating that correlation energy is not yet fully recovered. We attribute this residual error primarily to the use of CCSD for total energy calculations and expect that the inclusion of higher-order excitations, such as CCSDT, or the use of full configuration interaction would further improve accuracy. Basis-set extrapolation, discussed in Supplementary Information, may also contribute to the remaining discrepancy, although its effect is expected to be minor.

\section*{Conclusion}
In this work, we present a framework for constructing accurate many-body Hamiltonians for molecular systems using LCCVOs. By systematically improving the representation of virtual orbital spaces, the LCCVO approach overcomes a key limitation of conventional electronic-structure schemes, in which inadequately described unoccupied states lead to poor correlation treatments. We demonstrate that, with a substantially reduced number of orbitals, the resulting many-body Hamiltonians yield dissociation energies for both closed- and open-shell molecules that are comparable to, or even exceed, those obtained using large correlation-consistent basis sets. These results highlight the efficiency, scalability, and robustness of the LCCVO framework, offering a practical route toward high-accuracy quantum chemical calculations with reduced computational cost. Extension of this approach to periodic systems with explicit k-point sampling will be an important direction for future work. 

\section*{Methods}

\subsection*{Plane-Wave Many-Body Hamiltonian and Second Quantization}
Motivated by Bloch’s theorem, PWs can be seen as a grid basis in momentum space \cite{grosso2013solid}. By expanding the one-electron KS wavefunctions $\phi_{\sigma,n\mathbf{k}}(x)$ in terms of Bloch states, the molecular orbitals are represented as
\begin{equation}
\phi_{\sigma,n\mathbf{k}}(x) = \frac{e^{i\mathbf{k}\cdot\mathbf{x}}}{\sqrt{\Omega}}\sum_{\mathbf{G}}c_{\sigma,n\mathbf{k}}(\mathbf{G})e^{i\mathbf{G}\cdot\mathbf{x}},
\end{equation}
where $\sigma \in \{\uparrow,\downarrow\}$ indicates the spin projection (spin-up or spin-down), $n$ the band index, $\mathbf{k}$ the wave vector, $\Omega$ the volume of the primitive cell, $\mathbf{G}$ the reciprocal lattice vectors, and $c_{\sigma,n\mathbf{k}}(\mathbf{G})$ the coefficients of the PWs describing the wavefunction. As we focus on molecular systems ($\mathbf{k} = \Gamma$) in this study, we omit $\mathbf{k}$ hereafter.

The PW basis set is limited to a sphere in reciprocal space, with radius $|G_{\text{max}}|$ satisfying $\frac{1}{2}|G_{\text{max}}|^2 \leq E_{\text{cut}}$ where $E_{\text{cut}}$ is the cut-off energy. This framework enables granular control over the basis-set size via $E_{\text{cut}}$, allowing systematic convergence of the ground-state energy. In contrast, Gaussian-type orbitals offer less flexibility, as their basis-set expansion relies on discrete function forms with fixed increments in basis-set sizes.

Given a set of molecular orbitals for each of the studied species, the electronic Hamiltonian in second-quantized form is described as the following
\begin{equation}
H = \sum_{pq}^{M}h_{pq}a_p^{\dagger}a_q + \frac{1}{2}\sum_{pqrs}^{M}h_{pqrs}a_p^{\dagger}a_q^{\dagger}a_ra_s.
\end{equation}
Here, $h_{pq}$ represents the one-electron integrals, which has contributions from the electron’s kinetic energies and the electron-ion pseudopotential energies, while $h_{pqrs}$ represents the two-electron integrals, which accounts for the electron-electron Coulombic interactions. Conventionally, the Hartree-Fock (HF) states are used as the basis to compute $h_{pq}$ and $h_{pqrs}$. Recently, the PW basis set has been used to construct the one-electron wavefunctions, on top of which the many-body Hamiltonian can be solved by methods such as FCI, CC, M$\o$ller-Plesset perturbation theory. While the PW basis set is mostly adopted for the periodic systems, it can also be useful for molecular systems due to the granular control of the basis-set error \cite{qiu2025improved} and lack of basis-set superposition error. Furthermore, Kohn-Sham (KS) states obtained via DFT calculations can also be used instead of the HF states, which may serve as a better single-particle basis to construct the many-body Hamiltonian due to the partial inclusion of electron correlation in the KS wavefunction.

\subsection*{Correlation-Optimized Virtual Orbitals and Localized Correlation Converged Virtual Orbitals}
Performing many-body calculations of isolated systems using the plane-wave basis requires additional considerations. Due the periodic nature of plane-waves, spurious interactions with the periodic images of the molecules may be inadvertently introduced. While this is typically remedied by including a large vacuum region between the periodic images, the uniform nature of the plane-waves treats the isolated molecule and the vacuum region on an equal footing. As a result, PW-based DFT calculations will yield unoccupied electronic states with significant contributions from the vacuum region, including pure vacuum states and mixed vacuum-molecule states (referred to as partial vacuum states hereafter). The inclusion of such vacuum states considerably reduces the correlation energies captured in many-body calculations.

To address this issue, the concept of COVOs was proposed. In the original COVO method, the KS virtual orbitals are initially sorted in ascending KS orbital energies. Using these as an initial guess, new virtual orbitals are then obtained by optimizing the correlation energies of a very small active space consisting of only the highest occupied orbital and the single target virtual orbital~\cite{song2023periodic}. Despite the use of a minuscule active space, the COVO method has proven its success in capturing significant correlation energies, although its accuracy is often sacrificed.

Another important issue in quantum molecular computation is the determination of active space. Due to limited computational resources, it is impossible to utilize a truly complete basis set. Therefore, one has to compromise between accuracy and computational efficiency. It is especially so when using PW-based DFT calculations where much more unoccupied virtual orbitals can be obtained compared to an equivalent calculation using an atom-centred basis-set. While simply sorting the virtual orbitals by their KS orbital energies has proved reasonably successful for the original COVO method~\cite{bylaska2021quantum}, virtual orbitals considered in this manner may not have significant correlation energies, making it an inaccurate and inefficient basis.

We introduced LCCVOs, derived from the same PW-based DFT calculations, to address these issues. As mentioned previously, the pure and partial vacuum states may cause problems during correlation energy optimization. Thus when constructing the LCCVOs, we include the KS states in an energy-ascending manner and discard those with insignificant contributions to the correlation energy obtained by a CCSD calculation.
We demonstrate the LCCVO algorithm on several molecules, including singlet, doublet, and triplet molecules such as \ce{H2}, \ce{O2}, \ce{CN}, \ce{N2}, and \ce{C2} (see Supplementary algorithms 1 and 2). To evaluate the dissociation energies, the LCCVO calculations are also performed on the H, C, O, and N atoms, and the resulting molecular dissociation energies are compared with experimental values to assess the accuracy of our method. The results reproduce the expected $\mathcal{O}(1/N^{\text{orb}})$ error scaling with respect to the number of orbitals $N^{\text{orb}}$ (see Supplementary Figure 5). Remarkably, our method reduces the number of required spatial orbitals from 110--182 (using cc-pV5Z) to only 15--50, while maintaining (and sometimes surpassing) the accuracy of the much larger basis sets. Therefore, out method allows for more flexibility in the choice of $N^{\text{orb}}$ and achieves comparable correlation energy recovery with substantially fewer orbitals.
\medskip

\subsection*{Computational Methods}
All the DFT calculation of the molecular systems are performed using the Quantum \textsc{ESPRESSO}~\cite{giannozzi2009quantum,giannozzi2017advanced} software package, using the norm-conserving ONCV~\cite{hamannOptimizedNormconservingVanderbilt2013} pseudopotential with a wavefunction cutoff of \qty{80}{Ry}.
The resulting Kohn-Sham wavefunctions are then used as an initial guess for the LCCVO method, implemented according to Supplementary Algorithms 1 and 2, with PySCF~\cite{sun2020recent,sun2018pyscf} used to perform the required CCSD calculations.
To account for finite size effects inherent in the plane-wave basis, each calculation is performed using a series of simple cubic simulation cells with edge lengths of $L = \qtylist{10;12;14;16;18}{\AA}$, and the final energy is obtained by extrapolating the results to $L=\infty$ (see Supplementary Figure 6).
We also benchmarked our results against CBS extrapolations derived from the cc-pVXZ (X = T, Q, 5)\cite{halkier1999basis, peterson1993benchmark, peterson1995intrinsic} bases. For these basis sets, CCSD calculations are performed on a classical computer using the PySCF code~\cite{sun2020recent, sun2018pyscf}.

\section*{Acknowledgements}
A.H. gratefully acknowledges the sponsorship from National Natural Science Foundation of China (NSFC) (Grant No. 62541160274), City University of Hong Kong (Project No. 7006103), CityU Seed Fund in Microelectronics (Project No. 9229135), and Hon Hai Research Institute (Project No. 9231594, 9239182).
This work was carried out using the computational facilities, CityU Burgundy, managed and provided by the Computing Services Centre at City University of Hong Kong (\url{https://www.cityu.edu.hk}).

\section*{Author contributions}
Q.W. and C.K. contributed equally to this work.
The project was conceived by P.C., J.C., A.H., and M.H.
Software development and numerical simulations were performed by Q.W. and C.K.
Analysis of the data was performed by Q.W., P.C., and J.C.
All authors participated in the discussion and approved the final manuscript.

\bibliography{references}

\begin{thebibliography}{31}%
\makeatletter
\providecommand \@ifxundefined [1]{%
 \@ifx{#1\undefined}
}%
\providecommand \@ifnum [1]{%
 \ifnum #1\expandafter \@firstoftwo
 \else \expandafter \@secondoftwo
 \fi
}%
\providecommand \@ifx [1]{%
 \ifx #1\expandafter \@firstoftwo
 \else \expandafter \@secondoftwo
 \fi
}%
\providecommand \natexlab [1]{#1}%
\providecommand \enquote  [1]{``#1''}%
\providecommand \bibnamefont  [1]{#1}%
\providecommand \bibfnamefont [1]{#1}%
\providecommand \citenamefont [1]{#1}%
\providecommand \href@noop [0]{\@secondoftwo}%
\providecommand \href [0]{\begingroup \@sanitize@url \@href}%
\providecommand \@href[1]{\@@startlink{#1}\@@href}%
\providecommand \@@href[1]{\endgroup#1\@@endlink}%
\providecommand \@sanitize@url [0]{\catcode `\\12\catcode `\$12\catcode `\&12\catcode `\#12\catcode `\^12\catcode `\_12\catcode `\%12\relax}%
\providecommand \@@startlink[1]{}%
\providecommand \@@endlink[0]{}%
\providecommand \url  [0]{\begingroup\@sanitize@url \@url }%
\providecommand \@url [1]{\endgroup\@href {#1}{\urlprefix }}%
\providecommand \urlprefix  [0]{URL }%
\providecommand \Eprint [0]{\href }%
\providecommand \doibase [0]{https://doi.org/}%
\providecommand \selectlanguage [0]{\@gobble}%
\providecommand \bibinfo  [0]{\@secondoftwo}%
\providecommand \bibfield  [0]{\@secondoftwo}%
\providecommand \translation [1]{[#1]}%
\providecommand \BibitemOpen [0]{}%
\providecommand \bibitemStop [0]{}%
\providecommand \bibitemNoStop [0]{.\EOS\space}%
\providecommand \EOS [0]{\spacefactor3000\relax}%
\providecommand \BibitemShut  [1]{\csname bibitem#1\endcsname}%
\let\auto@bib@innerbib\@empty
\bibitem [{\citenamefont {Abrams}\ and\ \citenamefont {Lloyd}(1997)}]{abrams1997simulation}%
  \BibitemOpen
  \bibfield  {author} {\bibinfo {author} {\bibfnamefont {D.~S.}\ \bibnamefont {Abrams}}\ and\ \bibinfo {author} {\bibfnamefont {S.}~\bibnamefont {Lloyd}},\ }\bibfield  {title} {\bibinfo {title} {Simulation of many-body fermi systems on a universal quantum computer},\ }\href {https://doi.org/10.1103/PhysRevLett.79.2586} {\bibfield  {journal} {\bibinfo  {journal} {Physical Review Letters}\ }\textbf {\bibinfo {volume} {79}},\ \bibinfo {pages} {2586} (\bibinfo {year} {1997})}\BibitemShut {NoStop}%
\bibitem [{\citenamefont {Ku}\ \emph {et~al.}(2026)\citenamefont {Ku}, \citenamefont {Chen}, \citenamefont {Hu},\ and\ \citenamefont {Hsieh}}]{kuOptimizingQuantumChemistry2026}%
  \BibitemOpen
  \bibfield  {author} {\bibinfo {author} {\bibfnamefont {C.}~\bibnamefont {Ku}}, \bibinfo {author} {\bibfnamefont {Y.-C.}\ \bibnamefont {Chen}}, \bibinfo {author} {\bibfnamefont {A.}~\bibnamefont {Hu}},\ and\ \bibinfo {author} {\bibfnamefont {M.-H.}\ \bibnamefont {Hsieh}},\ }\bibfield  {title} {\bibinfo {title} {Optimizing quantum chemistry simulations with a hybrid quantization scheme},\ }\bibfield  {journal} {\bibinfo  {journal} {Communications Physics}\ }\href {https://doi.org/10.1038/s42005-026-02577-9} {10.1038/s42005-026-02577-9} (\bibinfo {year} {2026})\BibitemShut {NoStop}%
\bibitem [{\citenamefont {Berry}\ \emph {et~al.}(2019)\citenamefont {Berry}, \citenamefont {Gidney}, \citenamefont {Motta}, \citenamefont {McClean},\ and\ \citenamefont {Babbush}}]{berryQubitizationArbitraryBasis2019}%
  \BibitemOpen
  \bibfield  {author} {\bibinfo {author} {\bibfnamefont {D.~W.}\ \bibnamefont {Berry}}, \bibinfo {author} {\bibfnamefont {C.}~\bibnamefont {Gidney}}, \bibinfo {author} {\bibfnamefont {M.}~\bibnamefont {Motta}}, \bibinfo {author} {\bibfnamefont {J.~R.}\ \bibnamefont {McClean}},\ and\ \bibinfo {author} {\bibfnamefont {R.}~\bibnamefont {Babbush}},\ }\bibfield  {title} {\bibinfo {title} {Qubitization of {{Arbitrary Basis Quantum Chemistry Leveraging Sparsity}} and {{Low Rank Factorization}}},\ }\href {https://doi.org/10.22331/q-2019-12-02-208} {\bibfield  {journal} {\bibinfo  {journal} {Quantum}\ }\textbf {\bibinfo {volume} {3}},\ \bibinfo {pages} {208} (\bibinfo {year} {2019})}\BibitemShut {NoStop}%
\bibitem [{\citenamefont {Von~Burg}\ \emph {et~al.}(2021)\citenamefont {Von~Burg}, \citenamefont {Low}, \citenamefont {H{\"a}ner}, \citenamefont {Steiger}, \citenamefont {Reiher}, \citenamefont {Roetteler},\ and\ \citenamefont {Troyer}}]{vonburgQuantumComputingEnhanced2021}%
  \BibitemOpen
  \bibfield  {author} {\bibinfo {author} {\bibfnamefont {V.}~\bibnamefont {Von~Burg}}, \bibinfo {author} {\bibfnamefont {G.~H.}\ \bibnamefont {Low}}, \bibinfo {author} {\bibfnamefont {T.}~\bibnamefont {H{\"a}ner}}, \bibinfo {author} {\bibfnamefont {D.~S.}\ \bibnamefont {Steiger}}, \bibinfo {author} {\bibfnamefont {M.}~\bibnamefont {Reiher}}, \bibinfo {author} {\bibfnamefont {M.}~\bibnamefont {Roetteler}},\ and\ \bibinfo {author} {\bibfnamefont {M.}~\bibnamefont {Troyer}},\ }\bibfield  {title} {\bibinfo {title} {Quantum computing enhanced computational catalysis},\ }\href {https://doi.org/10.1103/PhysRevResearch.3.033055} {\bibfield  {journal} {\bibinfo  {journal} {Physical Review Research}\ }\textbf {\bibinfo {volume} {3}},\ \bibinfo {pages} {033055} (\bibinfo {year} {2021})}\BibitemShut {NoStop}%
\bibitem [{\citenamefont {Lee}\ \emph {et~al.}(2021)\citenamefont {Lee}, \citenamefont {Berry}, \citenamefont {Gidney}, \citenamefont {Huggins}, \citenamefont {McClean}, \citenamefont {Wiebe},\ and\ \citenamefont {Babbush}}]{leeEvenMoreEfficient2021}%
  \BibitemOpen
  \bibfield  {author} {\bibinfo {author} {\bibfnamefont {J.}~\bibnamefont {Lee}}, \bibinfo {author} {\bibfnamefont {D.~W.}\ \bibnamefont {Berry}}, \bibinfo {author} {\bibfnamefont {C.}~\bibnamefont {Gidney}}, \bibinfo {author} {\bibfnamefont {W.~J.}\ \bibnamefont {Huggins}}, \bibinfo {author} {\bibfnamefont {J.~R.}\ \bibnamefont {McClean}}, \bibinfo {author} {\bibfnamefont {N.}~\bibnamefont {Wiebe}},\ and\ \bibinfo {author} {\bibfnamefont {R.}~\bibnamefont {Babbush}},\ }\bibfield  {title} {\bibinfo {title} {Even {{More Efficient Quantum Computations}} of {{Chemistry Through Tensor Hypercontraction}}},\ }\href {https://doi.org/10.1103/PRXQuantum.2.030305} {\bibfield  {journal} {\bibinfo  {journal} {PRX Quantum}\ }\textbf {\bibinfo {volume} {2}},\ \bibinfo {pages} {030305} (\bibinfo {year} {2021})}\BibitemShut {NoStop}%
\bibitem [{\citenamefont {Georges}\ \emph {et~al.}(2025)\citenamefont {Georges}, \citenamefont {Bothe}, \citenamefont {S{\"u}nderhauf}, \citenamefont {Berntson}, \citenamefont {Izs{\'a}k},\ and\ \citenamefont {Ivanov}}]{georgesQuantumSimulationsChemistry2025}%
  \BibitemOpen
  \bibfield  {author} {\bibinfo {author} {\bibfnamefont {T.~N.}\ \bibnamefont {Georges}}, \bibinfo {author} {\bibfnamefont {M.}~\bibnamefont {Bothe}}, \bibinfo {author} {\bibfnamefont {C.}~\bibnamefont {S{\"u}nderhauf}}, \bibinfo {author} {\bibfnamefont {B.~K.}\ \bibnamefont {Berntson}}, \bibinfo {author} {\bibfnamefont {R.}~\bibnamefont {Izs{\'a}k}},\ and\ \bibinfo {author} {\bibfnamefont {A.~V.}\ \bibnamefont {Ivanov}},\ }\bibfield  {title} {\bibinfo {title} {Quantum simulations of chemistry in first quantization with any basis set},\ }\href {https://doi.org/10.1038/s41534-025-00987-1} {\bibfield  {journal} {\bibinfo  {journal} {npj Quantum Information}\ }\textbf {\bibinfo {volume} {11}},\ \bibinfo {pages} {55} (\bibinfo {year} {2025})}\BibitemShut {NoStop}%
\bibitem [{\citenamefont {Bauer}\ \emph {et~al.}(2020)\citenamefont {Bauer}, \citenamefont {Bravyi}, \citenamefont {Motta},\ and\ \citenamefont {Chan}}]{bauer2020quantum}%
  \BibitemOpen
  \bibfield  {author} {\bibinfo {author} {\bibfnamefont {B.}~\bibnamefont {Bauer}}, \bibinfo {author} {\bibfnamefont {S.}~\bibnamefont {Bravyi}}, \bibinfo {author} {\bibfnamefont {M.}~\bibnamefont {Motta}},\ and\ \bibinfo {author} {\bibfnamefont {G.~K.-L.}\ \bibnamefont {Chan}},\ }\bibfield  {title} {\bibinfo {title} {Quantum algorithms for quantum chemistry and quantum materials science},\ }\href {https://doi.org/10.1021/acs.chemrev.9b00829} {\bibfield  {journal} {\bibinfo  {journal} {Chemical Reviews}\ }\textbf {\bibinfo {volume} {120}},\ \bibinfo {pages} {12685} (\bibinfo {year} {2020})}\BibitemShut {NoStop}%
\bibitem [{\citenamefont {Preskill}(2018)}]{preskill2018quantum}%
  \BibitemOpen
  \bibfield  {author} {\bibinfo {author} {\bibfnamefont {J.}~\bibnamefont {Preskill}},\ }\bibfield  {title} {\bibinfo {title} {Quantum computing in the {{NISQ}} era and beyond},\ }\href {https://doi.org/10.22331/q-2018-08-06-79} {\bibfield  {journal} {\bibinfo  {journal} {Quantum}\ }\textbf {\bibinfo {volume} {2}},\ \bibinfo {pages} {79} (\bibinfo {year} {2018})}\BibitemShut {NoStop}%
\bibitem [{\citenamefont {Dunning}(1989)}]{dunning1989gaussian}%
  \BibitemOpen
  \bibfield  {author} {\bibinfo {author} {\bibfnamefont {T.~H.}\ \bibnamefont {Dunning}, \bibfnamefont {Jr.}},\ }\bibfield  {title} {\bibinfo {title} {Gaussian basis sets for use in correlated molecular calculations. {{I}}. {{The}} atoms boron through neon and hydrogen},\ }\href {https://doi.org/10.1063/1.456153} {\bibfield  {journal} {\bibinfo  {journal} {Journal of Chemical Physics}\ }\textbf {\bibinfo {volume} {90}},\ \bibinfo {pages} {1007} (\bibinfo {year} {1989})}\BibitemShut {NoStop}%
\bibitem [{\citenamefont {Song}\ \emph {et~al.}(2023)\citenamefont {Song}, \citenamefont {Bauman}, \citenamefont {Prawiroatmodjo}, \citenamefont {Peng}, \citenamefont {Granade}, \citenamefont {Rosso}, \citenamefont {Low}, \citenamefont {Roetteler}, \citenamefont {Kowalski},\ and\ \citenamefont {Bylaska}}]{song2023periodic}%
  \BibitemOpen
  \bibfield  {author} {\bibinfo {author} {\bibfnamefont {D.}~\bibnamefont {Song}}, \bibinfo {author} {\bibfnamefont {N.~P.}\ \bibnamefont {Bauman}}, \bibinfo {author} {\bibfnamefont {G.}~\bibnamefont {Prawiroatmodjo}}, \bibinfo {author} {\bibfnamefont {B.}~\bibnamefont {Peng}}, \bibinfo {author} {\bibfnamefont {C.}~\bibnamefont {Granade}}, \bibinfo {author} {\bibfnamefont {K.~M.}\ \bibnamefont {Rosso}}, \bibinfo {author} {\bibfnamefont {G.~H.}\ \bibnamefont {Low}}, \bibinfo {author} {\bibfnamefont {M.}~\bibnamefont {Roetteler}}, \bibinfo {author} {\bibfnamefont {K.}~\bibnamefont {Kowalski}},\ and\ \bibinfo {author} {\bibfnamefont {E.~J.}\ \bibnamefont {Bylaska}},\ }\bibfield  {title} {\bibinfo {title} {Periodic plane-wave electronic structure calculations on quantum computers},\ }\href {https://doi.org/10.1186/s41313-022-00049-5} {\bibfield  {journal} {\bibinfo  {journal} {Materials Theory}\ }\textbf {\bibinfo {volume} {7}},\ \bibinfo {pages} {2} (\bibinfo {year} {2023})}\BibitemShut {NoStop}%
\bibitem [{\citenamefont {Bylaska}\ \emph {et~al.}(2021)\citenamefont {Bylaska}, \citenamefont {Song}, \citenamefont {Bauman}, \citenamefont {Kowalski}, \citenamefont {Claudino},\ and\ \citenamefont {Humble}}]{bylaska2021quantum}%
  \BibitemOpen
  \bibfield  {author} {\bibinfo {author} {\bibfnamefont {E.~J.}\ \bibnamefont {Bylaska}}, \bibinfo {author} {\bibfnamefont {D.}~\bibnamefont {Song}}, \bibinfo {author} {\bibfnamefont {N.~P.}\ \bibnamefont {Bauman}}, \bibinfo {author} {\bibfnamefont {K.}~\bibnamefont {Kowalski}}, \bibinfo {author} {\bibfnamefont {D.}~\bibnamefont {Claudino}},\ and\ \bibinfo {author} {\bibfnamefont {T.~S.}\ \bibnamefont {Humble}},\ }\bibfield  {title} {\bibinfo {title} {Quantum solvers for plane-wave hamiltonians: Abridging virtual spaces through the optimization of pairwise correlations},\ }\href {https://doi.org/10.3389/fchem.2021.603019} {\bibfield  {journal} {\bibinfo  {journal} {Frontiers in Chemistry}\ }\textbf {\bibinfo {volume} {9}},\ \bibinfo {pages} {603019} (\bibinfo {year} {2021})}\BibitemShut {NoStop}%
\bibitem [{\citenamefont {Herzberg}(1970)}]{herzberg1970dissociation}%
  \BibitemOpen
  \bibfield  {author} {\bibinfo {author} {\bibfnamefont {G.}~\bibnamefont {Herzberg}},\ }\bibfield  {title} {\bibinfo {title} {The dissociation energy of the hydrogen molecule},\ }\href {https://doi.org/10.1016/0022-2852(70)90060-3} {\bibfield  {journal} {\bibinfo  {journal} {Journal of Molecular Spectroscopy}\ }\textbf {\bibinfo {volume} {33}},\ \bibinfo {pages} {147} (\bibinfo {year} {1970})}\BibitemShut {NoStop}%
\bibitem [{\citenamefont {Lu}\ \emph {et~al.}(2023)\citenamefont {Lu}, \citenamefont {Jiang},\ and\ \citenamefont {Gao}}]{lu2023observation}%
  \BibitemOpen
  \bibfield  {author} {\bibinfo {author} {\bibfnamefont {L.}~\bibnamefont {Lu}}, \bibinfo {author} {\bibfnamefont {P.}~\bibnamefont {Jiang}},\ and\ \bibinfo {author} {\bibfnamefont {H.}~\bibnamefont {Gao}},\ }\bibfield  {title} {\bibinfo {title} {Observation of continuum state dissociation enables the determination of {{N}}{\textsubscript{2}} bond dissociation energy to spectroscopic accuracy},\ }\href {https://doi.org/10.1021/acs.jpclett.3c02665} {\bibfield  {journal} {\bibinfo  {journal} {Journal of Physical Chemistry Letters}\ }\textbf {\bibinfo {volume} {14}},\ \bibinfo {pages} {10974} (\bibinfo {year} {2023})}\BibitemShut {NoStop}%
\bibitem [{\citenamefont {Bytautas}\ and\ \citenamefont {Ruedenberg}(2005)}]{bytautas2005correlation}%
  \BibitemOpen
  \bibfield  {author} {\bibinfo {author} {\bibfnamefont {L.}~\bibnamefont {Bytautas}}\ and\ \bibinfo {author} {\bibfnamefont {K.}~\bibnamefont {Ruedenberg}},\ }\bibfield  {title} {\bibinfo {title} {Correlation energy extrapolation by intrinsic scaling. {{IV}}. {{Accurate}} binding energies of the homonuclear diatomic molecules carbon, nitrogen, oxygen, and fluorine},\ }\href {https://doi.org/10.1063/1.1869493} {\bibfield  {journal} {\bibinfo  {journal} {Journal of Chemical Physics}\ }\textbf {\bibinfo {volume} {122}},\ \bibinfo {pages} {154110} (\bibinfo {year} {2005})}\BibitemShut {NoStop}%
\bibitem [{\citenamefont {Wang}\ \emph {et~al.}(2024)\citenamefont {Wang}, \citenamefont {Gong},\ and\ \citenamefont {Mo}}]{wang2024bond}%
  \BibitemOpen
  \bibfield  {author} {\bibinfo {author} {\bibfnamefont {P.}~\bibnamefont {Wang}}, \bibinfo {author} {\bibfnamefont {S.}~\bibnamefont {Gong}},\ and\ \bibinfo {author} {\bibfnamefont {Y.}~\bibnamefont {Mo}},\ }\bibfield  {title} {\bibinfo {title} {Bond dissociation energy of {{O}}{\textsubscript{2}} measured by fully state-to-state resolved threshold fragment yield spectra},\ }\href {https://doi.org/10.1063/5.0207288} {\bibfield  {journal} {\bibinfo  {journal} {Journal of Chemical Physics}\ }\textbf {\bibinfo {volume} {160}},\ \bibinfo {pages} {164302} (\bibinfo {year} {2024})}\BibitemShut {NoStop}%
\bibitem [{\citenamefont {Pradhan}\ \emph {et~al.}(1994)\citenamefont {Pradhan}, \citenamefont {Partridge},\ and\ \citenamefont {Bauschlicher}}]{pradhan1994dissociation}%
  \BibitemOpen
  \bibfield  {author} {\bibinfo {author} {\bibfnamefont {A.~D.}\ \bibnamefont {Pradhan}}, \bibinfo {author} {\bibfnamefont {H.}~\bibnamefont {Partridge}},\ and\ \bibinfo {author} {\bibfnamefont {C.~W.}\ \bibnamefont {Bauschlicher}},\ }\bibfield  {title} {\bibinfo {title} {The dissociation energy of {{CN}} and {{C}}{\textsubscript{2}}},\ }\href {https://doi.org/10.1063/1.467503} {\bibfield  {journal} {\bibinfo  {journal} {Journal of Chemical Physics}\ }\textbf {\bibinfo {volume} {101}},\ \bibinfo {pages} {3857} (\bibinfo {year} {1994})}\BibitemShut {NoStop}%
\bibitem [{\citenamefont {Huang}\ \emph {et~al.}(1992)\citenamefont {Huang}, \citenamefont {Barts},\ and\ \citenamefont {Halpern}}]{huang1992heat}%
  \BibitemOpen
  \bibfield  {author} {\bibinfo {author} {\bibfnamefont {Y.}~\bibnamefont {Huang}}, \bibinfo {author} {\bibfnamefont {S.~A.}\ \bibnamefont {Barts}},\ and\ \bibinfo {author} {\bibfnamefont {J.~B.}\ \bibnamefont {Halpern}},\ }\bibfield  {title} {\bibinfo {title} {Heat of formation of the cyanogen radical},\ }\href {https://doi.org/10.1021/j100180a079} {\bibfield  {journal} {\bibinfo  {journal} {Journal of Physical Chemistry}\ }\textbf {\bibinfo {volume} {96}},\ \bibinfo {pages} {425} (\bibinfo {year} {1992})}\BibitemShut {NoStop}%
\bibitem [{\citenamefont {Su}\ \emph {et~al.}(2011)\citenamefont {Su}, \citenamefont {Wu}, \citenamefont {Gu}, \citenamefont {Wu}, \citenamefont {Shaik},\ and\ \citenamefont {Hiberty}}]{su2011bonding}%
  \BibitemOpen
  \bibfield  {author} {\bibinfo {author} {\bibfnamefont {P.}~\bibnamefont {Su}}, \bibinfo {author} {\bibfnamefont {J.}~\bibnamefont {Wu}}, \bibinfo {author} {\bibfnamefont {J.}~\bibnamefont {Gu}}, \bibinfo {author} {\bibfnamefont {W.}~\bibnamefont {Wu}}, \bibinfo {author} {\bibfnamefont {S.}~\bibnamefont {Shaik}},\ and\ \bibinfo {author} {\bibfnamefont {P.~C.}\ \bibnamefont {Hiberty}},\ }\bibfield  {title} {\bibinfo {title} {Bonding conundrums in the {{C}}{\textsubscript{2}} molecule: A valence bond study},\ }\href {https://doi.org/10.1021/ct100577v} {\bibfield  {journal} {\bibinfo  {journal} {Journal of Chemical Theory and Computation}\ }\textbf {\bibinfo {volume} {7}},\ \bibinfo {pages} {121} (\bibinfo {year} {2011})}\BibitemShut {NoStop}%
\bibitem [{\citenamefont {Martin}\ and\ \citenamefont {Hepburn}(1997)}]{martin1997electric}%
  \BibitemOpen
  \bibfield  {author} {\bibinfo {author} {\bibfnamefont {J.~D.~D.}\ \bibnamefont {Martin}}\ and\ \bibinfo {author} {\bibfnamefont {J.~W.}\ \bibnamefont {Hepburn}},\ }\bibfield  {title} {\bibinfo {title} {Electric field induced dissociation of molecules in rydberg-like highly vibrationally excited ion-pair states},\ }\href {https://doi.org/10.1103/PhysRevLett.79.3154} {\bibfield  {journal} {\bibinfo  {journal} {Physical Review Letters}\ }\textbf {\bibinfo {volume} {79}},\ \bibinfo {pages} {3154} (\bibinfo {year} {1997})}\BibitemShut {NoStop}%
\bibitem [{\citenamefont {Cohen}\ \emph {et~al.}(2008)\citenamefont {Cohen}, \citenamefont {{Mori-S{\'a}nchez}},\ and\ \citenamefont {Yang}}]{cohen2008fractional}%
  \BibitemOpen
  \bibfield  {author} {\bibinfo {author} {\bibfnamefont {A.~J.}\ \bibnamefont {Cohen}}, \bibinfo {author} {\bibfnamefont {P.}~\bibnamefont {{Mori-S{\'a}nchez}}},\ and\ \bibinfo {author} {\bibfnamefont {W.}~\bibnamefont {Yang}},\ }\bibfield  {title} {\bibinfo {title} {Fractional spins and static correlation error in density functional theory},\ }\href {https://doi.org/10.1063/1.2987202} {\bibfield  {journal} {\bibinfo  {journal} {Journal of Chemical Physics}\ }\textbf {\bibinfo {volume} {129}},\ \bibinfo {pages} {121104} (\bibinfo {year} {2008})}\BibitemShut {NoStop}%
\bibitem [{\citenamefont {Medvedev}\ \emph {et~al.}(2017)\citenamefont {Medvedev}, \citenamefont {Bushmarinov}, \citenamefont {Sun}, \citenamefont {Perdew},\ and\ \citenamefont {Lyssenko}}]{medvedev2017density}%
  \BibitemOpen
  \bibfield  {author} {\bibinfo {author} {\bibfnamefont {M.~G.}\ \bibnamefont {Medvedev}}, \bibinfo {author} {\bibfnamefont {I.~S.}\ \bibnamefont {Bushmarinov}}, \bibinfo {author} {\bibfnamefont {J.}~\bibnamefont {Sun}}, \bibinfo {author} {\bibfnamefont {J.~P.}\ \bibnamefont {Perdew}},\ and\ \bibinfo {author} {\bibfnamefont {K.~A.}\ \bibnamefont {Lyssenko}},\ }\bibfield  {title} {\bibinfo {title} {Density functional theory is straying from the path toward the exact functional},\ }\href {https://doi.org/10.1126/science.aah5975} {\bibfield  {journal} {\bibinfo  {journal} {Science}\ }\textbf {\bibinfo {volume} {355}},\ \bibinfo {pages} {49} (\bibinfo {year} {2017})}\BibitemShut {NoStop}%
\bibitem [{\citenamefont {Grosso}\ and\ \citenamefont {Pastori~Parravicini}(2000)}]{grosso2013solid}%
  \BibitemOpen
  \bibfield  {author} {\bibinfo {author} {\bibfnamefont {G.}~\bibnamefont {Grosso}}\ and\ \bibinfo {author} {\bibfnamefont {G.}~\bibnamefont {Pastori~Parravicini}},\ }\href@noop {} {\emph {\bibinfo {title} {Solid State Physics}}}\ (\bibinfo  {publisher} {Academic Press},\ \bibinfo {address} {San Diego},\ \bibinfo {year} {2000})\BibitemShut {NoStop}%
\bibitem [{\citenamefont {Qiu}\ \emph {et~al.}(2025)\citenamefont {Qiu}, \citenamefont {Zhang}, \citenamefont {Zhang}, \citenamefont {Lin}, \citenamefont {Li}, \citenamefont {Yang},\ and\ \citenamefont {Hu}}]{qiu2025improved}%
  \BibitemOpen
  \bibfield  {author} {\bibinfo {author} {\bibfnamefont {M.}~\bibnamefont {Qiu}}, \bibinfo {author} {\bibfnamefont {Z.}~\bibnamefont {Zhang}}, \bibinfo {author} {\bibfnamefont {Z.}~\bibnamefont {Zhang}}, \bibinfo {author} {\bibfnamefont {Y.}~\bibnamefont {Lin}}, \bibinfo {author} {\bibfnamefont {Y.}~\bibnamefont {Li}}, \bibinfo {author} {\bibfnamefont {J.}~\bibnamefont {Yang}},\ and\ \bibinfo {author} {\bibfnamefont {W.}~\bibnamefont {Hu}},\ }\bibfield  {title} {\bibinfo {title} {Improved correlation optimized virtual orbital algorithm for plane-wave full configuration interaction calculations},\ }\href {https://doi.org/10.1021/acs.jctc.5c00586} {\bibfield  {journal} {\bibinfo  {journal} {Journal of Chemical Theory and Computation}\ }\textbf {\bibinfo {volume} {21}},\ \bibinfo {pages} {6559} (\bibinfo {year} {2025})}\BibitemShut {NoStop}%
\bibitem [{\citenamefont {Giannozzi}\ \emph {et~al.}(2009)\citenamefont {Giannozzi}, \citenamefont {Baroni}, \citenamefont {Bonini}, \citenamefont {Calandra}, \citenamefont {Car}, \citenamefont {Cavazzoni}, \citenamefont {Ceresoli}, \citenamefont {Chiarotti}, \citenamefont {Cococcioni}, \citenamefont {Dabo}, \citenamefont {Corso}, \citenamefont {de~Gironcoli}, \citenamefont {Fabris}, \citenamefont {Fratesi}, \citenamefont {Gebauer}, \citenamefont {Gerstmann}, \citenamefont {Gougoussis}, \citenamefont {Kokalj}, \citenamefont {Lazzeri}, \citenamefont {{Martin-Samos}}, \citenamefont {Marzari}, \citenamefont {Mauri}, \citenamefont {Mazzarello}, \citenamefont {Paolini}, \citenamefont {Pasquarello}, \citenamefont {Paulatto}, \citenamefont {Sbraccia}, \citenamefont {Scandolo}, \citenamefont {Sclauzero}, \citenamefont {Seitsonen}, \citenamefont {Smogunov}, \citenamefont {Umari},\ and\ \citenamefont {Wentzcovitch}}]{giannozzi2009quantum}%
  \BibitemOpen
  \bibfield  {author} {\bibinfo {author} {\bibfnamefont {P.}~\bibnamefont {Giannozzi}}, \bibinfo {author} {\bibfnamefont {S.}~\bibnamefont {Baroni}}, \bibinfo {author} {\bibfnamefont {N.}~\bibnamefont {Bonini}}, \bibinfo {author} {\bibfnamefont {M.}~\bibnamefont {Calandra}}, \bibinfo {author} {\bibfnamefont {R.}~\bibnamefont {Car}}, \bibinfo {author} {\bibfnamefont {C.}~\bibnamefont {Cavazzoni}}, \bibinfo {author} {\bibfnamefont {D.}~\bibnamefont {Ceresoli}}, \bibinfo {author} {\bibfnamefont {G.~L.}\ \bibnamefont {Chiarotti}}, \bibinfo {author} {\bibfnamefont {M.}~\bibnamefont {Cococcioni}}, \bibinfo {author} {\bibfnamefont {I.}~\bibnamefont {Dabo}}, \bibinfo {author} {\bibfnamefont {A.~D.}\ \bibnamefont {Corso}}, \bibinfo {author} {\bibfnamefont {S.}~\bibnamefont {de~Gironcoli}}, \bibinfo {author} {\bibfnamefont {S.}~\bibnamefont {Fabris}}, \bibinfo {author} {\bibfnamefont {G.}~\bibnamefont {Fratesi}}, \bibinfo {author} {\bibfnamefont {R.}~\bibnamefont {Gebauer}}, \bibinfo {author} {\bibfnamefont
  {U.}~\bibnamefont {Gerstmann}}, \bibinfo {author} {\bibfnamefont {C.}~\bibnamefont {Gougoussis}}, \bibinfo {author} {\bibfnamefont {A.}~\bibnamefont {Kokalj}}, \bibinfo {author} {\bibfnamefont {M.}~\bibnamefont {Lazzeri}}, \bibinfo {author} {\bibfnamefont {L.}~\bibnamefont {{Martin-Samos}}}, \bibinfo {author} {\bibfnamefont {N.}~\bibnamefont {Marzari}}, \bibinfo {author} {\bibfnamefont {F.}~\bibnamefont {Mauri}}, \bibinfo {author} {\bibfnamefont {R.}~\bibnamefont {Mazzarello}}, \bibinfo {author} {\bibfnamefont {S.}~\bibnamefont {Paolini}}, \bibinfo {author} {\bibfnamefont {A.}~\bibnamefont {Pasquarello}}, \bibinfo {author} {\bibfnamefont {L.}~\bibnamefont {Paulatto}}, \bibinfo {author} {\bibfnamefont {C.}~\bibnamefont {Sbraccia}}, \bibinfo {author} {\bibfnamefont {S.}~\bibnamefont {Scandolo}}, \bibinfo {author} {\bibfnamefont {G.}~\bibnamefont {Sclauzero}}, \bibinfo {author} {\bibfnamefont {A.~P.}\ \bibnamefont {Seitsonen}}, \bibinfo {author} {\bibfnamefont {A.}~\bibnamefont {Smogunov}}, \bibinfo {author}
  {\bibfnamefont {P.}~\bibnamefont {Umari}},\ and\ \bibinfo {author} {\bibfnamefont {R.~M.}\ \bibnamefont {Wentzcovitch}},\ }\bibfield  {title} {\bibinfo {title} {{{QUANTUM ESPRESSO}}: A modular and open-source software project for quantum simulations of materials},\ }\href {https://doi.org/10.1088/0953-8984/21/39/395502} {\bibfield  {journal} {\bibinfo  {journal} {Journal of Physics: Condensed Matter}\ }\textbf {\bibinfo {volume} {21}},\ \bibinfo {pages} {395502} (\bibinfo {year} {2009})}\BibitemShut {NoStop}%
\bibitem [{\citenamefont {Giannozzi}\ \emph {et~al.}(2017)\citenamefont {Giannozzi}, \citenamefont {Andreussi}, \citenamefont {Brumme}, \citenamefont {Bunau}, \citenamefont {Nardelli}, \citenamefont {Calandra}, \citenamefont {Car}, \citenamefont {Cavazzoni}, \citenamefont {Ceresoli}, \citenamefont {Cococcioni}, \citenamefont {Colonna}, \citenamefont {Carnimeo}, \citenamefont {Corso}, \citenamefont {de~Gironcoli}, \citenamefont {Delugas}, \citenamefont {DiStasio}, \citenamefont {Ferretti}, \citenamefont {Floris}, \citenamefont {Fratesi}, \citenamefont {Fugallo}, \citenamefont {Gebauer}, \citenamefont {Gerstmann}, \citenamefont {Giustino}, \citenamefont {Gorni}, \citenamefont {Jia}, \citenamefont {Kawamura}, \citenamefont {Ko}, \citenamefont {Kokalj}, \citenamefont {K{\"u}{\c c}{\"u}kbenli}, \citenamefont {Lazzeri}, \citenamefont {Marsili}, \citenamefont {Marzari}, \citenamefont {Mauri}, \citenamefont {Nguyen}, \citenamefont {Nguyen}, \citenamefont {{Otero-de-la-Roza}}, \citenamefont {Paulatto}, \citenamefont
  {Ponc{\'e}}, \citenamefont {Rocca}, \citenamefont {Sabatini}, \citenamefont {Santra}, \citenamefont {Schlipf}, \citenamefont {Seitsonen}, \citenamefont {Smogunov}, \citenamefont {Timrov}, \citenamefont {Thonhauser}, \citenamefont {Umari}, \citenamefont {Vast}, \citenamefont {Wu},\ and\ \citenamefont {Baroni}}]{giannozzi2017advanced}%
  \BibitemOpen
  \bibfield  {author} {\bibinfo {author} {\bibfnamefont {P.}~\bibnamefont {Giannozzi}}, \bibinfo {author} {\bibfnamefont {O.}~\bibnamefont {Andreussi}}, \bibinfo {author} {\bibfnamefont {T.}~\bibnamefont {Brumme}}, \bibinfo {author} {\bibfnamefont {O.}~\bibnamefont {Bunau}}, \bibinfo {author} {\bibfnamefont {M.~B.}\ \bibnamefont {Nardelli}}, \bibinfo {author} {\bibfnamefont {M.}~\bibnamefont {Calandra}}, \bibinfo {author} {\bibfnamefont {R.}~\bibnamefont {Car}}, \bibinfo {author} {\bibfnamefont {C.}~\bibnamefont {Cavazzoni}}, \bibinfo {author} {\bibfnamefont {D.}~\bibnamefont {Ceresoli}}, \bibinfo {author} {\bibfnamefont {M.}~\bibnamefont {Cococcioni}}, \bibinfo {author} {\bibfnamefont {N.}~\bibnamefont {Colonna}}, \bibinfo {author} {\bibfnamefont {I.}~\bibnamefont {Carnimeo}}, \bibinfo {author} {\bibfnamefont {A.~D.}\ \bibnamefont {Corso}}, \bibinfo {author} {\bibfnamefont {S.}~\bibnamefont {de~Gironcoli}}, \bibinfo {author} {\bibfnamefont {P.}~\bibnamefont {Delugas}}, \bibinfo {author} {\bibfnamefont
  {R.~A.}\ \bibnamefont {DiStasio}}, \bibinfo {author} {\bibfnamefont {A.}~\bibnamefont {Ferretti}}, \bibinfo {author} {\bibfnamefont {A.}~\bibnamefont {Floris}}, \bibinfo {author} {\bibfnamefont {G.}~\bibnamefont {Fratesi}}, \bibinfo {author} {\bibfnamefont {G.}~\bibnamefont {Fugallo}}, \bibinfo {author} {\bibfnamefont {R.}~\bibnamefont {Gebauer}}, \bibinfo {author} {\bibfnamefont {U.}~\bibnamefont {Gerstmann}}, \bibinfo {author} {\bibfnamefont {F.}~\bibnamefont {Giustino}}, \bibinfo {author} {\bibfnamefont {T.}~\bibnamefont {Gorni}}, \bibinfo {author} {\bibfnamefont {J.}~\bibnamefont {Jia}}, \bibinfo {author} {\bibfnamefont {M.}~\bibnamefont {Kawamura}}, \bibinfo {author} {\bibfnamefont {H.-Y.}\ \bibnamefont {Ko}}, \bibinfo {author} {\bibfnamefont {A.}~\bibnamefont {Kokalj}}, \bibinfo {author} {\bibfnamefont {E.}~\bibnamefont {K{\"u}{\c c}{\"u}kbenli}}, \bibinfo {author} {\bibfnamefont {M.}~\bibnamefont {Lazzeri}}, \bibinfo {author} {\bibfnamefont {M.}~\bibnamefont {Marsili}}, \bibinfo {author}
  {\bibfnamefont {N.}~\bibnamefont {Marzari}}, \bibinfo {author} {\bibfnamefont {F.}~\bibnamefont {Mauri}}, \bibinfo {author} {\bibfnamefont {N.~L.}\ \bibnamefont {Nguyen}}, \bibinfo {author} {\bibfnamefont {H.-V.}\ \bibnamefont {Nguyen}}, \bibinfo {author} {\bibfnamefont {A.}~\bibnamefont {{Otero-de-la-Roza}}}, \bibinfo {author} {\bibfnamefont {L.}~\bibnamefont {Paulatto}}, \bibinfo {author} {\bibfnamefont {S.}~\bibnamefont {Ponc{\'e}}}, \bibinfo {author} {\bibfnamefont {D.}~\bibnamefont {Rocca}}, \bibinfo {author} {\bibfnamefont {R.}~\bibnamefont {Sabatini}}, \bibinfo {author} {\bibfnamefont {B.}~\bibnamefont {Santra}}, \bibinfo {author} {\bibfnamefont {M.}~\bibnamefont {Schlipf}}, \bibinfo {author} {\bibfnamefont {A.~P.}\ \bibnamefont {Seitsonen}}, \bibinfo {author} {\bibfnamefont {A.}~\bibnamefont {Smogunov}}, \bibinfo {author} {\bibfnamefont {I.}~\bibnamefont {Timrov}}, \bibinfo {author} {\bibfnamefont {T.}~\bibnamefont {Thonhauser}}, \bibinfo {author} {\bibfnamefont {P.}~\bibnamefont {Umari}}, \bibinfo
  {author} {\bibfnamefont {N.}~\bibnamefont {Vast}}, \bibinfo {author} {\bibfnamefont {X.}~\bibnamefont {Wu}},\ and\ \bibinfo {author} {\bibfnamefont {S.}~\bibnamefont {Baroni}},\ }\bibfield  {title} {\bibinfo {title} {Advanced capabilities for materials modelling with {{Quantum ESPRESSO}}},\ }\href {https://doi.org/10.1088/1361-648X/aa8f79} {\bibfield  {journal} {\bibinfo  {journal} {Journal of Physics: Condensed Matter}\ }\textbf {\bibinfo {volume} {29}},\ \bibinfo {pages} {465901} (\bibinfo {year} {2017})}\BibitemShut {NoStop}%
\bibitem [{\citenamefont {Hamann}(2013)}]{hamannOptimizedNormconservingVanderbilt2013}%
  \BibitemOpen
  \bibfield  {author} {\bibinfo {author} {\bibfnamefont {D.~R.}\ \bibnamefont {Hamann}},\ }\bibfield  {title} {\bibinfo {title} {Optimized norm-conserving {{Vanderbilt}} pseudopotentials},\ }\href {https://doi.org/10.1103/PhysRevB.88.085117} {\bibfield  {journal} {\bibinfo  {journal} {Physical Review B}\ }\textbf {\bibinfo {volume} {88}},\ \bibinfo {pages} {085117} (\bibinfo {year} {2013})}\BibitemShut {NoStop}%
\bibitem [{\citenamefont {Sun}\ \emph {et~al.}(2020)\citenamefont {Sun}, \citenamefont {Zhang}, \citenamefont {Banerjee}, \citenamefont {Bao}, \citenamefont {Barbry}, \citenamefont {Blunt}, \citenamefont {Bogdanov}, \citenamefont {Booth}, \citenamefont {Chen}, \citenamefont {Cui}, \citenamefont {Eriksen}, \citenamefont {Gao}, \citenamefont {Guo}, \citenamefont {Hermann}, \citenamefont {Hermes}, \citenamefont {Koh}, \citenamefont {Koval}, \citenamefont {Lehtola}, \citenamefont {Li}, \citenamefont {Liu}, \citenamefont {Mardirossian}, \citenamefont {McClain}, \citenamefont {Motta}, \citenamefont {Mussard}, \citenamefont {Pham}, \citenamefont {Pulkin}, \citenamefont {Purwanto}, \citenamefont {Robinson}, \citenamefont {Ronca}, \citenamefont {Sayfutyarova}, \citenamefont {Scheurer}, \citenamefont {Schurkus}, \citenamefont {Smith}, \citenamefont {Sun}, \citenamefont {Sun}, \citenamefont {Upadhyay}, \citenamefont {Wagner}, \citenamefont {Wang}, \citenamefont {White}, \citenamefont {Whitfield}, \citenamefont
  {Williamson}, \citenamefont {Wouters}, \citenamefont {Yang}, \citenamefont {Yu}, \citenamefont {Zhu}, \citenamefont {Berkelbach}, \citenamefont {Sharma}, \citenamefont {Sokolov},\ and\ \citenamefont {Chan}}]{sun2020recent}%
  \BibitemOpen
  \bibfield  {author} {\bibinfo {author} {\bibfnamefont {Q.}~\bibnamefont {Sun}}, \bibinfo {author} {\bibfnamefont {X.}~\bibnamefont {Zhang}}, \bibinfo {author} {\bibfnamefont {S.}~\bibnamefont {Banerjee}}, \bibinfo {author} {\bibfnamefont {P.}~\bibnamefont {Bao}}, \bibinfo {author} {\bibfnamefont {M.}~\bibnamefont {Barbry}}, \bibinfo {author} {\bibfnamefont {N.~S.}\ \bibnamefont {Blunt}}, \bibinfo {author} {\bibfnamefont {N.~A.}\ \bibnamefont {Bogdanov}}, \bibinfo {author} {\bibfnamefont {G.~H.}\ \bibnamefont {Booth}}, \bibinfo {author} {\bibfnamefont {J.}~\bibnamefont {Chen}}, \bibinfo {author} {\bibfnamefont {Z.-H.}\ \bibnamefont {Cui}}, \bibinfo {author} {\bibfnamefont {J.~J.}\ \bibnamefont {Eriksen}}, \bibinfo {author} {\bibfnamefont {Y.}~\bibnamefont {Gao}}, \bibinfo {author} {\bibfnamefont {S.}~\bibnamefont {Guo}}, \bibinfo {author} {\bibfnamefont {J.}~\bibnamefont {Hermann}}, \bibinfo {author} {\bibfnamefont {M.~R.}\ \bibnamefont {Hermes}}, \bibinfo {author} {\bibfnamefont {K.}~\bibnamefont {Koh}},
  \bibinfo {author} {\bibfnamefont {P.}~\bibnamefont {Koval}}, \bibinfo {author} {\bibfnamefont {S.}~\bibnamefont {Lehtola}}, \bibinfo {author} {\bibfnamefont {Z.}~\bibnamefont {Li}}, \bibinfo {author} {\bibfnamefont {J.}~\bibnamefont {Liu}}, \bibinfo {author} {\bibfnamefont {N.}~\bibnamefont {Mardirossian}}, \bibinfo {author} {\bibfnamefont {J.~D.}\ \bibnamefont {McClain}}, \bibinfo {author} {\bibfnamefont {M.}~\bibnamefont {Motta}}, \bibinfo {author} {\bibfnamefont {B.}~\bibnamefont {Mussard}}, \bibinfo {author} {\bibfnamefont {H.~Q.}\ \bibnamefont {Pham}}, \bibinfo {author} {\bibfnamefont {A.}~\bibnamefont {Pulkin}}, \bibinfo {author} {\bibfnamefont {W.}~\bibnamefont {Purwanto}}, \bibinfo {author} {\bibfnamefont {P.~J.}\ \bibnamefont {Robinson}}, \bibinfo {author} {\bibfnamefont {E.}~\bibnamefont {Ronca}}, \bibinfo {author} {\bibfnamefont {E.~R.}\ \bibnamefont {Sayfutyarova}}, \bibinfo {author} {\bibfnamefont {M.}~\bibnamefont {Scheurer}}, \bibinfo {author} {\bibfnamefont {H.~F.}\ \bibnamefont {Schurkus}},
  \bibinfo {author} {\bibfnamefont {J.~E.~T.}\ \bibnamefont {Smith}}, \bibinfo {author} {\bibfnamefont {C.}~\bibnamefont {Sun}}, \bibinfo {author} {\bibfnamefont {S.-N.}\ \bibnamefont {Sun}}, \bibinfo {author} {\bibfnamefont {S.}~\bibnamefont {Upadhyay}}, \bibinfo {author} {\bibfnamefont {L.~K.}\ \bibnamefont {Wagner}}, \bibinfo {author} {\bibfnamefont {X.}~\bibnamefont {Wang}}, \bibinfo {author} {\bibfnamefont {A.}~\bibnamefont {White}}, \bibinfo {author} {\bibfnamefont {J.~D.}\ \bibnamefont {Whitfield}}, \bibinfo {author} {\bibfnamefont {M.~J.}\ \bibnamefont {Williamson}}, \bibinfo {author} {\bibfnamefont {S.}~\bibnamefont {Wouters}}, \bibinfo {author} {\bibfnamefont {J.}~\bibnamefont {Yang}}, \bibinfo {author} {\bibfnamefont {J.~M.}\ \bibnamefont {Yu}}, \bibinfo {author} {\bibfnamefont {T.}~\bibnamefont {Zhu}}, \bibinfo {author} {\bibfnamefont {T.~C.}\ \bibnamefont {Berkelbach}}, \bibinfo {author} {\bibfnamefont {S.}~\bibnamefont {Sharma}}, \bibinfo {author} {\bibfnamefont {A.~Y.}\ \bibnamefont
  {Sokolov}},\ and\ \bibinfo {author} {\bibfnamefont {G.~K.-L.}\ \bibnamefont {Chan}},\ }\bibfield  {title} {\bibinfo {title} {Recent developments in the {{PySCF}} program package},\ }\href {https://doi.org/10.1063/5.0006074} {\bibfield  {journal} {\bibinfo  {journal} {The Journal of Chemical Physics}\ }\textbf {\bibinfo {volume} {153}},\ \bibinfo {pages} {024109} (\bibinfo {year} {2020})}\BibitemShut {NoStop}%
\bibitem [{\citenamefont {Sun}\ \emph {et~al.}(2018)\citenamefont {Sun}, \citenamefont {Berkelbach}, \citenamefont {Blunt}, \citenamefont {Booth}, \citenamefont {Guo}, \citenamefont {Li}, \citenamefont {Liu}, \citenamefont {McClain}, \citenamefont {Sayfutyarova}, \citenamefont {Sharma}, \citenamefont {Wouters},\ and\ \citenamefont {Chan}}]{sun2018pyscf}%
  \BibitemOpen
  \bibfield  {author} {\bibinfo {author} {\bibfnamefont {Q.}~\bibnamefont {Sun}}, \bibinfo {author} {\bibfnamefont {T.~C.}\ \bibnamefont {Berkelbach}}, \bibinfo {author} {\bibfnamefont {N.~S.}\ \bibnamefont {Blunt}}, \bibinfo {author} {\bibfnamefont {G.~H.}\ \bibnamefont {Booth}}, \bibinfo {author} {\bibfnamefont {S.}~\bibnamefont {Guo}}, \bibinfo {author} {\bibfnamefont {Z.}~\bibnamefont {Li}}, \bibinfo {author} {\bibfnamefont {J.}~\bibnamefont {Liu}}, \bibinfo {author} {\bibfnamefont {J.~D.}\ \bibnamefont {McClain}}, \bibinfo {author} {\bibfnamefont {E.~R.}\ \bibnamefont {Sayfutyarova}}, \bibinfo {author} {\bibfnamefont {S.}~\bibnamefont {Sharma}}, \bibinfo {author} {\bibfnamefont {S.}~\bibnamefont {Wouters}},\ and\ \bibinfo {author} {\bibfnamefont {G.~K.-L.}\ \bibnamefont {Chan}},\ }\bibfield  {title} {\bibinfo {title} {P{\textsc{y}}{{SCF}}: The python-based simulations of chemistry framework},\ }\href {https://doi.org/10.1002/wcms.1340} {\bibfield  {journal} {\bibinfo  {journal} {WIREs Computational
  Molecular Science}\ }\textbf {\bibinfo {volume} {8}},\ \bibinfo {pages} {e1340} (\bibinfo {year} {2018})}\BibitemShut {NoStop}%
\bibitem [{\citenamefont {Halkier}\ \emph {et~al.}(1999)\citenamefont {Halkier}, \citenamefont {Helgaker}, \citenamefont {J{\o}rgensen}, \citenamefont {Klopper},\ and\ \citenamefont {Olsen}}]{halkier1999basis}%
  \BibitemOpen
  \bibfield  {author} {\bibinfo {author} {\bibfnamefont {A.}~\bibnamefont {Halkier}}, \bibinfo {author} {\bibfnamefont {T.}~\bibnamefont {Helgaker}}, \bibinfo {author} {\bibfnamefont {P.}~\bibnamefont {J{\o}rgensen}}, \bibinfo {author} {\bibfnamefont {W.}~\bibnamefont {Klopper}},\ and\ \bibinfo {author} {\bibfnamefont {J.}~\bibnamefont {Olsen}},\ }\bibfield  {title} {\bibinfo {title} {Basis-set convergence of the energy in molecular {{Hartree}}--{{Fock}} calculations},\ }\href {https://doi.org/10.1016/S0009-2614(99)00179-7} {\bibfield  {journal} {\bibinfo  {journal} {Chemical Physics Letters}\ }\textbf {\bibinfo {volume} {302}},\ \bibinfo {pages} {437} (\bibinfo {year} {1999})}\BibitemShut {NoStop}%
\bibitem [{\citenamefont {Peterson}\ \emph {et~al.}(1993)\citenamefont {Peterson}, \citenamefont {Kendall},\ and\ \citenamefont {Dunning}}]{peterson1993benchmark}%
  \BibitemOpen
  \bibfield  {author} {\bibinfo {author} {\bibfnamefont {K.~A.}\ \bibnamefont {Peterson}}, \bibinfo {author} {\bibfnamefont {R.~A.}\ \bibnamefont {Kendall}},\ and\ \bibinfo {author} {\bibfnamefont {T.~H.}\ \bibnamefont {Dunning}},\ }\bibfield  {title} {\bibinfo {title} {Benchmark calculations with correlated molecular wave functions. {{II}}. {{Configuration}} interaction calculations on first row diatomic hydrides},\ }\href {https://doi.org/10.1063/1.465307} {\bibfield  {journal} {\bibinfo  {journal} {The Journal of Chemical Physics}\ }\textbf {\bibinfo {volume} {99}},\ \bibinfo {pages} {1930} (\bibinfo {year} {1993})}\BibitemShut {NoStop}%
\bibitem [{\citenamefont {Peterson}\ and\ \citenamefont {Dunning}(1995)}]{peterson1995intrinsic}%
  \BibitemOpen
  \bibfield  {author} {\bibinfo {author} {\bibfnamefont {K.~A.}\ \bibnamefont {Peterson}}\ and\ \bibinfo {author} {\bibfnamefont {T.~H.}\ \bibnamefont {Dunning}},\ }\bibfield  {title} {\bibinfo {title} {Intrinsic errors in several ab initio methods: The dissociation energy of {{N2}}},\ }\href {https://doi.org/10.1021/j100012a005} {\bibfield  {journal} {\bibinfo  {journal} {The Journal of Physical Chemistry}\ }\textbf {\bibinfo {volume} {99}},\ \bibinfo {pages} {3898} (\bibinfo {year} {1995})}\BibitemShut {NoStop}%
\end{thebibliography}%
\clearpage

\onecolumngrid

\renewcommand{\figurename}{Supplementary Figure}
\floatname{algorithm}{Supplementary Algorithm}

\makeatletter
\renewcommand{\section}{\@startsection{section}{1}{0pt}{-\baselineskip}{0.5\baselineskip}{\raggedright\bfseries}}
\makeatother

\setcounter{page}{1}
\setcounter{figure}{0}
\setcounter{table}{0}
\setcounter{equation}{0}
\setcounter{algorithm}{0}

\section*{Supplementary Note 1: Remarks on the Dissociation Energy of \ce{C2}}
As mentioned in the main text, the error in the dissociation energy of \ce{C2} exceeds 10\% when using the proposed PW-derived LCCVO approach. Although this deviation remains unsatisfactory in absolute terms, it is important to emphasize that the present method still yields the smallest error among all approaches considered in this study. Since the same framework provides reliable results for open-shell systems such as triplet \ce{O2} and the doublet CN radical, the open-shell character of \ce{C2} is unlikely to be the dominant source of error.

Instead, the discrepancy may be partly attributed to the uncertainty associated with experimental methods. In practice, \ce{C2} is typically generated under conditions where weak interactions with surrounding species or transient environments cannot be completely eliminated. Even subtle perturbations may influence its geometric and electronic structures, which in turn affect the experimentally inferred dissociation energy.
Additional ambiguity is reflected in the determination of the ground-state electronic structure of \ce{C2}. Although it is commonly assigned as a singlet X$^1\Sigma_{g}^{+}$ state in experiments, its characterization is known to be highly sensitive, suggesting that subtle factors may influence the interpretation of experimental observations.

Taken together, these considerations indicate that the relatively large error in the dissociation energy is unlikely to arise solely from the LCCVO method itself, but may also be related to the intrinsic difficulty in establishing a definitive experimental reference for \ce{C2}. Therefore, the present results should be interpreted in the context of this inherent uncertainty.

\section*{Supplementary Figures}

\begin{itemize}
  \item Supplementary Figure 1: The total energies as a function of interatomic distance using LCCVO for singlet state systems, along with comparisons of dissociation energies with atom-centred bases.
  \item Supplementary Figure 2: The total energies as a function of interatomic distance using LCCVO for doublet state systems, along with comparisons of dissociation energies with atom-centred bases.
  \item Supplementary Figure 3: The total energies as a function of interatomic distance using LCCVO for triplet state systems, along with comparisons of dissociation energies with atom-centred bases.
  \item Supplementary Figure 4: Bar chart comparing the correlation energies obtained with and without LCCVO.
  \item Supplementary Figure 5: Plot of CCSD energies of the \ce{N2} system with varying numbers of LCCVOs, illustrating the $\mathcal{O}(1/N)$ relationship between CCSD energies and the number of LCCVOs.
  \item Supplementary Figure 6: Plot of CCSD energies of the \ce{N2} system with varying cell parameters, illustrating the $\mathcal{O}(1/L)$ relationship between the CCSD energies and the cell parameters.
\end{itemize}

\clearpage

\begin{figure}[H]
  \centering
  \includegraphics[width=1\textwidth]{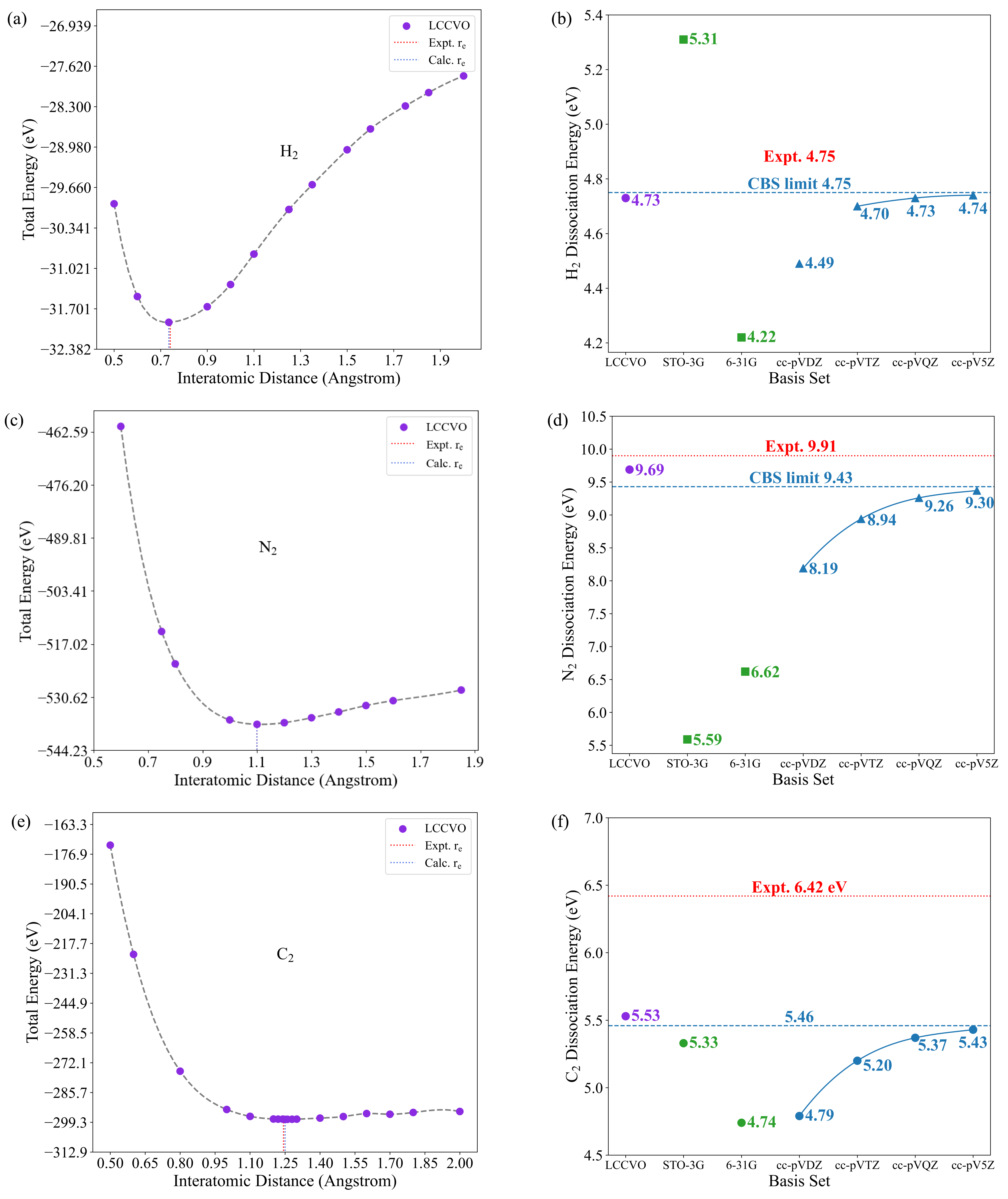}
  \caption{
    The total energies as a function of interatomic distance using LCCVO for singlet state systems, along with comparisons of dissociation energies with atom-centred bases.
    (a), (c), and (e) shows the LCCVO total energies of \ce{H2}, \ce{N2}, and \ce{C2}, respectively.
    (b), (d), and (f) shows the LCCVO dissociation energies $D_e$ at the equilibrium bond lengths, comparing it against various atom-centred basis-sets for \ce{H2}, \ce{N2}, and \ce{C2}, respectively.
    The blue dashed line shows the CBS limit value of the cc-pVXZ basis-sets.
  }
  \label{fig:supp1}
\end{figure}

\begin{figure}[H]
  \centering
  \includegraphics[width=1\textwidth]{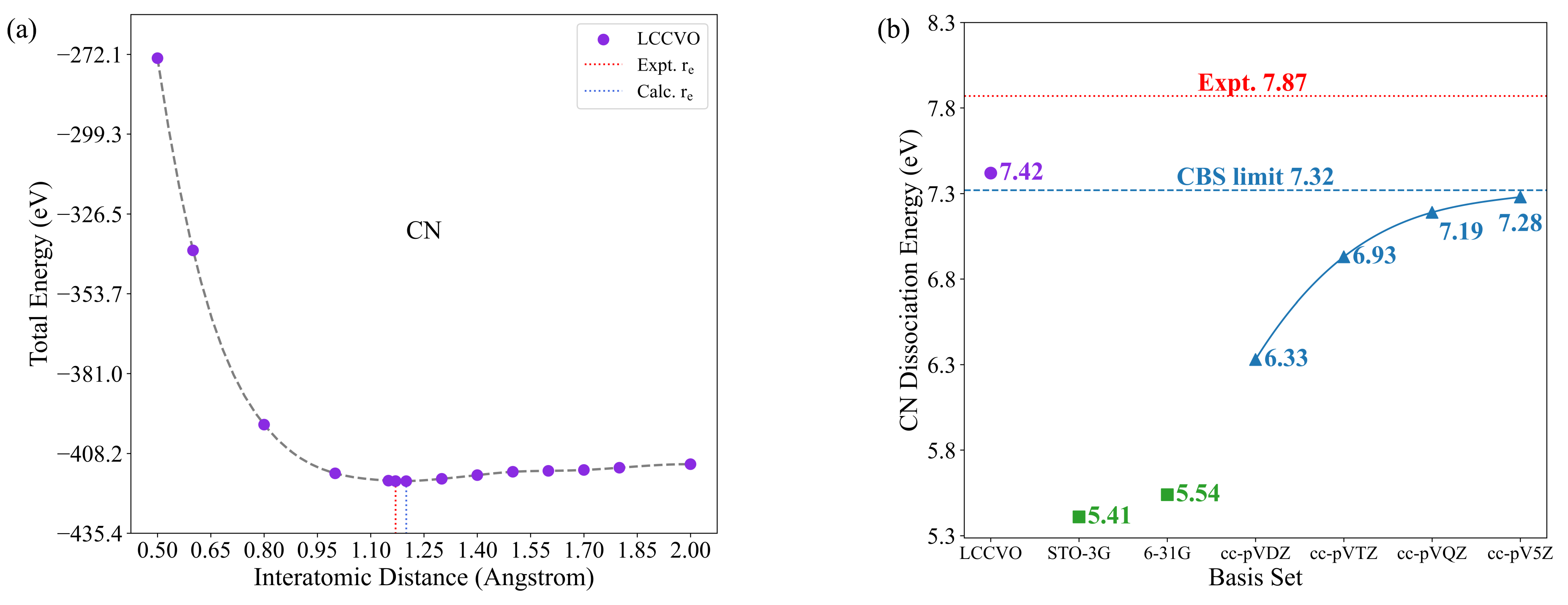}
  \caption{
    The total energies as a function of interatomic distance using LCCVO for doublet state systems, along with comparisons of dissociation energies with atom-centred bases.
    (a) shows the LCCVO total energies of \ce{CN}.
    (b) shows the LCCVO dissociation energies $D_e$ at the equilibrium bond lengths, comparing it against various atom-centred basis-sets for \ce{CN}.
    The blue dashed line shows the extrapolated energies of the cc-pVXZ basis-sets.
  }
  \label{fig:supp2}
\end{figure}

\begin{figure}[H]
  \centering
  \includegraphics[width=1\textwidth]{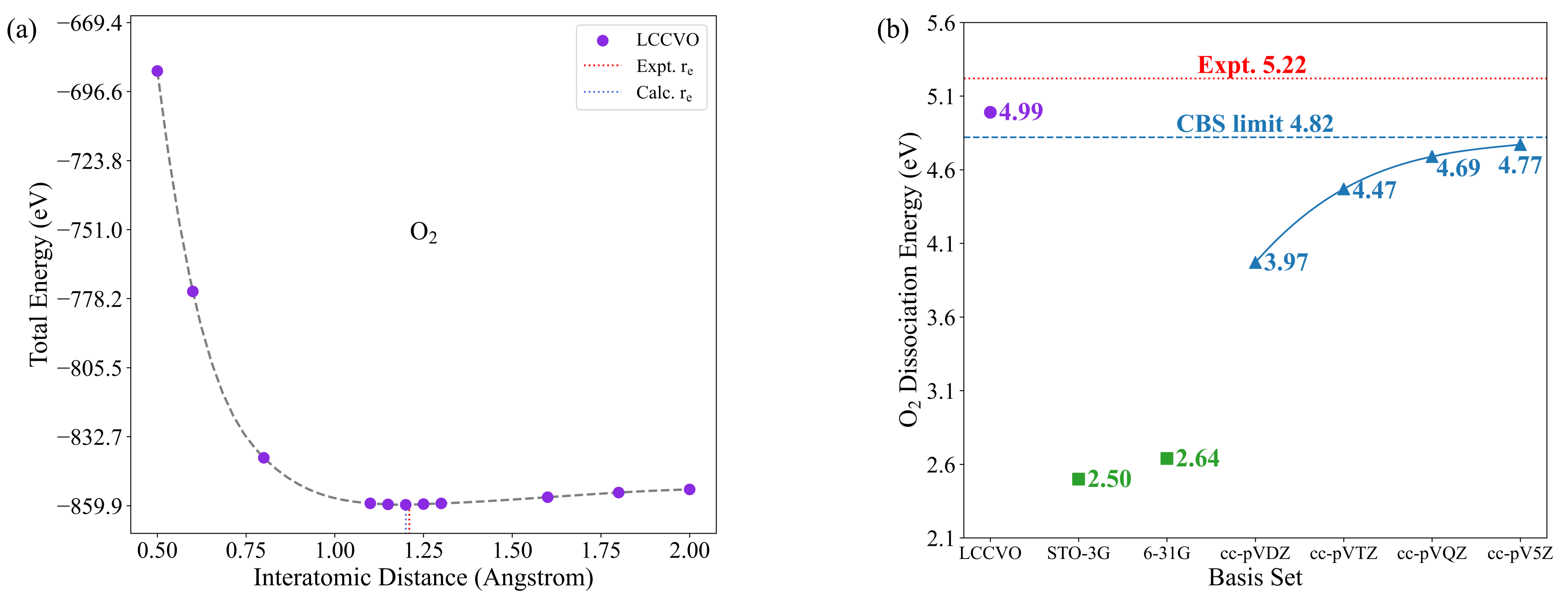}
  \caption{
    The total energies as a function of interatomic distance using LCCVO for triplet state systems, along with comparisons of dissociation energies with atom-centred bases.
    (a) shows the LCCVO total energies of \ce{O2}.
    (b) shows the LCCVO dissociation energies $D_e$ at the equilibrium bond lengths, comparing it against various atom-centred basis-sets for \ce{O2}.
    The blue dashed line shows the extrapolated energies of the cc-pVXZ basis-sets.
  }
  \label{fig:supp3}
\end{figure}

\clearpage

\begin{figure}[H]
  \centering
  \includegraphics[width=1\textwidth]{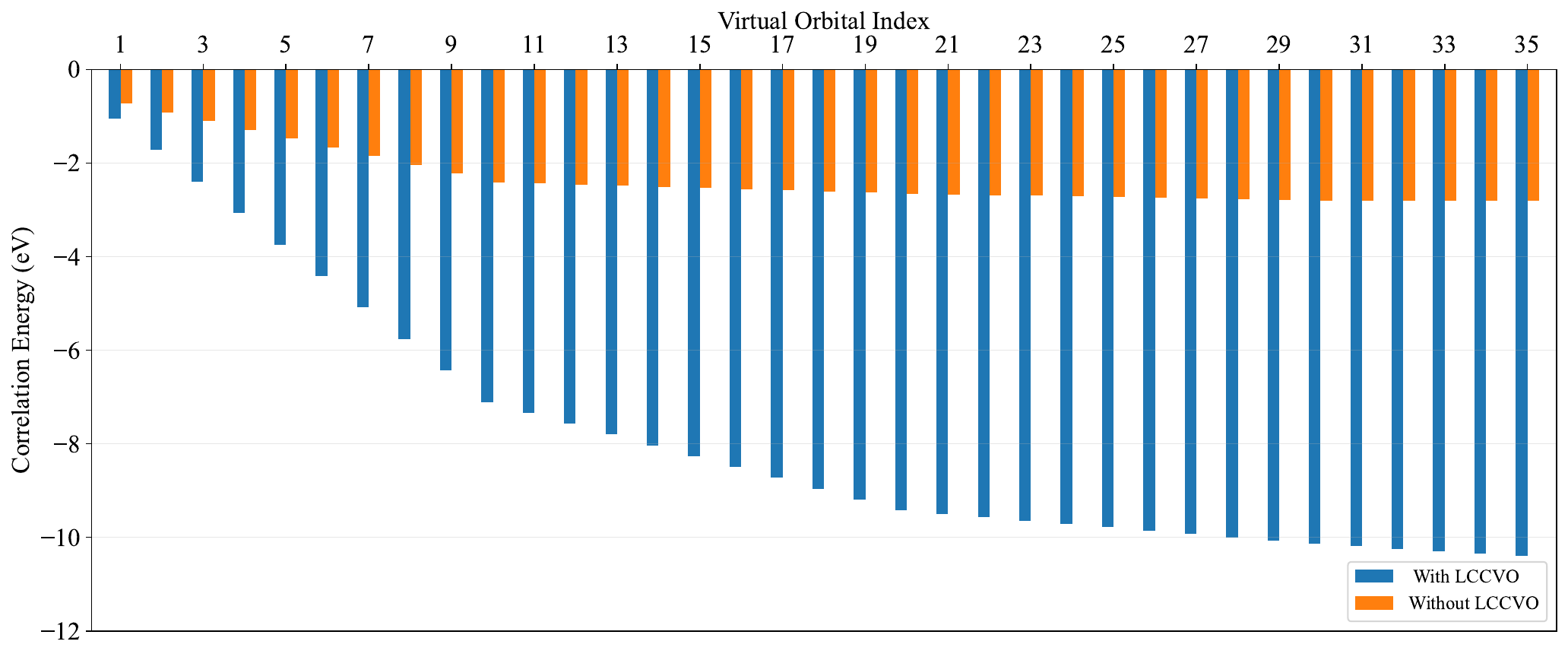}
  \caption{
    Bar chart comparing the correlation energies obtained with and without LCCVO.
    This was performed for the \ce{N2} system using 1--35 virtual orbitals at its equilibrium bond lengths of \qty{1.10}{\AA} using a cubic simulation cell of \qty{16}{\AA} side length.
  }
  \label{fig:supp4}
\end{figure}

\begin{figure}[H]
  \centering
  \includegraphics[width=0.8\textwidth]{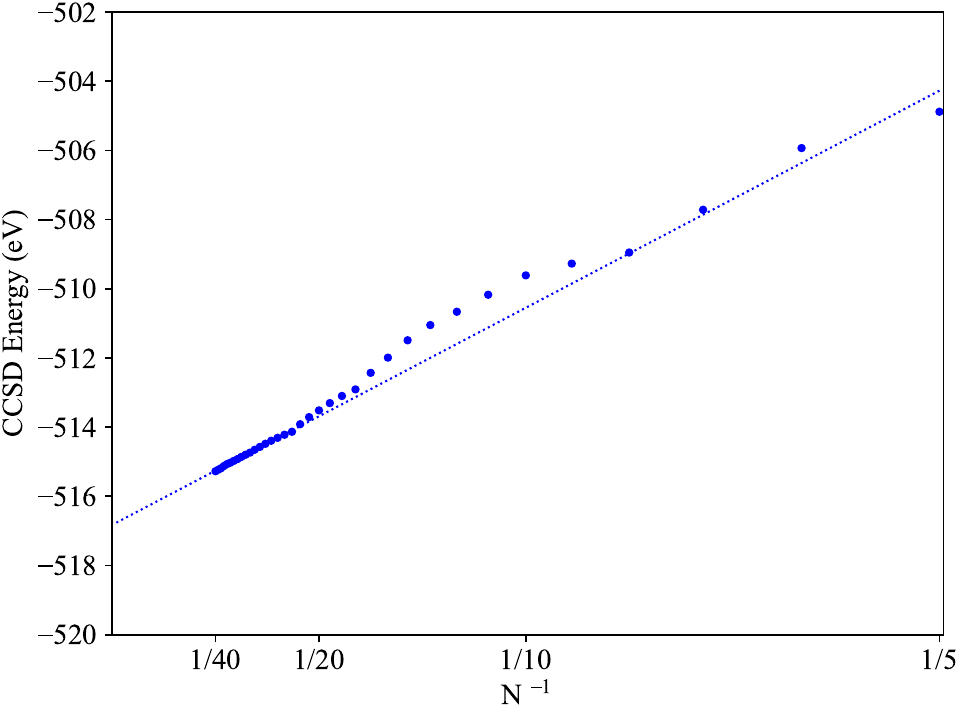}
  \caption{
    Plot of CCSD energies of the \ce{N2} system with varying numbers of LCCVOs, illustrating the $\mathcal{O}(1/N)$ relationship between CCSD energies and the number of LCCVOs.
    The $y$-axis represents the CCSD energies, while the $x$-axis is the reciprocal of the number of spatial orbitals.
    The dotted line shows the $\mathcal{O}(1/N)$ energy scaling for calculations larger than \num{16} LCCVOs.
  }
  \label{fig:supp5}
\end{figure}

\clearpage

\begin{figure}[H]
  \centering
  \includegraphics[width=0.8\textwidth]{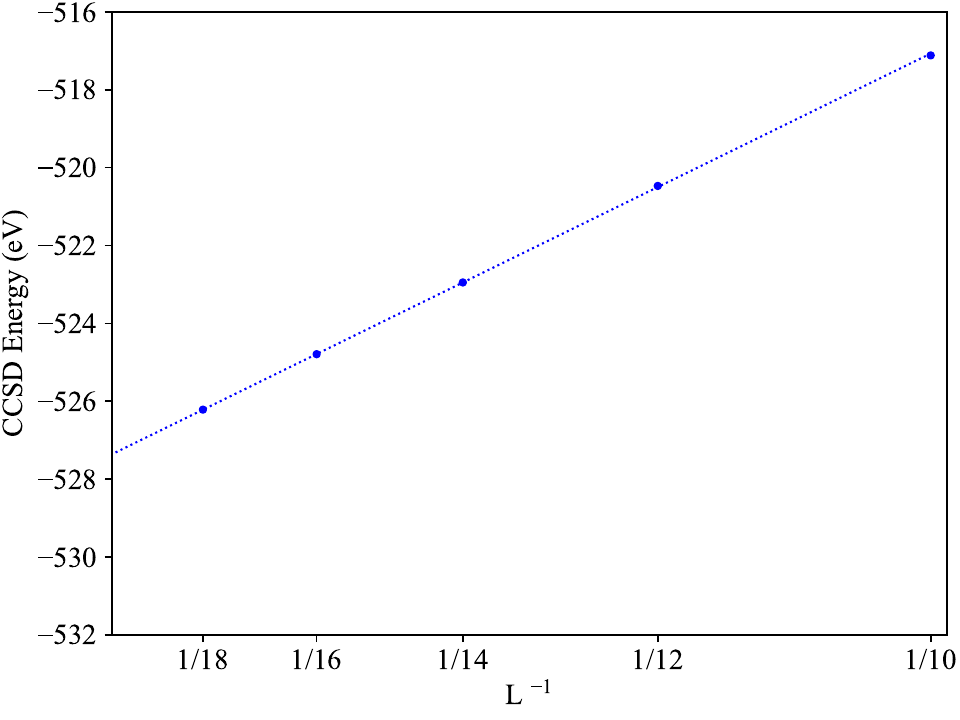}
  \caption{
    Plot of CCSD energies of the \ce{N2} system with varying cell parameters, illustrating the $\mathcal{O}(1/L)$ relationship between the CCSD energies and the cell parameters.
    The $y$-axis represents the CCSD energies, while the $x$-axis is the reciprocal of the cell parameters.
    The dotted line shows the extrapolation process that take advantage of this energy scaling, using the energy at $L=\infty$.
  }
  \label{fig:supp6}
\end{figure}

\clearpage

\section*{Supplementary Algorithms}

\begin{itemize}

  \item Supplementary Algorithm 1: Pseudocode to obtain a single LCCVO $\hat{\psi}_{N+t}$ for the singlet state system.
  
  \item Supplementary Algorithm 2: Pseudocode to obtain a single LCCVO $\hat{\psi}_{(\sigma,N+t)}$ ($\sigma = \uparrow, \downarrow$) for the doublet and triplet state system.

\end{itemize}

\setcounter{algorithm}{0}
\begin{algorithm}[H]
  \caption{Pseudocode to obtain a single LCCVO $\hat{\psi}_{N+t}$ for the singlet state system.}
  \label{alg:covo-single-virtual}
  \begin{algorithmic}[1]
    \State \textbf{Input:}
    \begin{itemize}[nosep]
      \item Initial virtual orbitals: $\psi_{N+t}$
      \item Occupied orbitals: $\psi_1, \psi_2, \ldots, \psi_{N-1}, \psi_N$
      \item Previously generated virtual orbitals: $\hat{\psi}_{N+1}, \hat{\psi}_{N+2}, \ldots, \hat{\psi}_{N+(t-1)}$
      \item Convergence thresholds: $\epsilon_1, \epsilon_2$
      \item Maximum iterations: $M$
    \end{itemize}
    \State \textbf{Output:} Optimized orbital $\hat{\psi}_{N+t}$
    \State Construct set $S$:
    \begin{equation*}
        S = \{\psi_1, \psi_2, \ldots, \psi_{N-1}, \psi_N, \hat{\psi}_{N+1}, \hat{\psi}_{N+2}, \ldots, \hat{\psi}_{N+(t-1)}\}
    \end{equation*}
    \State Initialize iteration counter $k \leftarrow 0$, error $\mathit{error}_1 \leftarrow \infty$, $\mathit{error}_2 \leftarrow \infty$, learning rate $\gamma \leftarrow \gamma_0$
    \While{$(\mathit{error}_1 > \epsilon_1 \ \mathbf{or} \ \mathit{error}_2 > \epsilon_2)$ \textbf{and} $k < M$} 
      \State Orthogonalize and normalize $\psi_{N+t}^k$ with respect to all orbitals in $S$
      \State Construct $H_{\mathit{CCSD}}$ for $2N$-electron system with $N + t$ spatial orbitals ($\psi_{N+t}$ and $S$)
      \State Compute gradient:
      \begin{equation*}
          \frac{\partial E_{\mathit{CCSD}}^k}{\partial \psi_{N+t}^k}
      \end{equation*}
      \State Update parameters:
      \begin{equation*}
        \psi_{N+t}^{k+1} \leftarrow \psi_{N+t}^k - \gamma \, \dfrac{\partial E_{\mathit{CCSD}}^k}{\partial \psi_{N+t}^k}
      \end{equation*}
      \State Compute new energy error: $\mathit{error}_1 \leftarrow E_{\mathit{CCSD}}^{k+1} - E_{\mathit{CCSD}}^k$
      \State Compute new wavefunction error: $\mathit{error}_2 \leftarrow \big\|\psi_{N+t}^{k+1} - \psi_{N+t}^k\big\|$
      \If{$E_{\mathit{CCSD}}^{k+1} > E_{\mathit{CCSD}}^k$}
        \State Reduce learning rate: $\gamma \leftarrow \gamma/2$
        \State Reset descent counter: $\mathit{consecutive\_descent} \leftarrow 0$
        \State \textbf{continue}
      \Else
        \State $\mathit{consecutive\_descent} \leftarrow \mathit{consecutive\_descent} + 1$
        \If{$\mathit{consecutive\_descent} > 5$}
          \State Slightly increase learning rate: $\gamma \leftarrow \gamma \times 1.1$
        \EndIf
      \EndIf
      \State Increment iteration counter: $k \leftarrow k + 1$
    \EndWhile
    \State Orthogonalize and normalize $\psi_{N+t}$ with respect to all orbitals in $S$
    \State \textbf{return} optimized orbital $\hat{\psi}_{N+t}$
  \end{algorithmic}
\end{algorithm}

\clearpage

\begin{algorithm}[H]
  \caption{Pseudocode to obtain a single LCCVO $\hat{\psi}_{(\sigma,N+t)}$ ($\sigma = \uparrow, \downarrow$) for the doublet and triplet state system. We assume $N+1$ spin-up occupied orbitals and $N-1$ spin-down occupied orbitals for this example.}
  \label{alg:covo-spin-polarized}
  \begin{algorithmic}[1]
    \State \textbf{Input:}
    \begin{itemize}[nosep]
      \item Initial virtual orbitals: $\psi_{(\uparrow, N+t)}, \psi_{(\downarrow, N+t)}$
      \item Occupied orbitals: $\psi_{(\uparrow,1)}, \psi_{(\downarrow,1)}, \psi_{(\uparrow,2)}, \psi_{(\downarrow,2)}, \ldots, \psi_{(\uparrow,N-1)}, \psi_{(\downarrow,N-1)}, \psi_{(\uparrow,N)},\psi_{(\uparrow,N+1)}$
      \item Previously generated virtual orbitals: $\hat{\psi}_{(\downarrow,N)}, \hat{\psi}_{(\downarrow,N+1)}, \hat{\psi}_{(\downarrow,N+2)}, \hat{\psi}_{(\downarrow,N+2)},\ldots, \hat{\psi}_{(\uparrow,N+t-1)}, \hat{\psi}_{(\downarrow,N+t-1)}$
      \item Convergence thresholds: $\epsilon_1, \epsilon_2$
      \item Maximum iterations: $M$
    \end{itemize}
    \State \textbf{Output:} Optimized orbitals $\hat{\psi}_{(\uparrow, N+t)}$ and $\hat{\psi}_{(\downarrow, N+t)}$
    \State Construct set $S$:
    \begin{align*}
        S = \{
        &\psi_{(\uparrow,1)}, \psi_{(\downarrow,1)}, \psi_{(\uparrow,2)}, \psi_{(\downarrow,2)},\ldots, \psi_{(\uparrow,N-1)}, \psi_{(\downarrow,N-1)}, \psi_{(\uparrow,N)}, \psi_{(\uparrow,N+1)},\\
        &\hat{\psi}_{(\downarrow,N)}, \hat{\psi}_{(\downarrow,N+1)},
        \hat{\psi}_{(\uparrow,N+2)}, \hat{\psi}_{(\downarrow,N+2)}, \ldots,
        \hat{\psi}_{(\uparrow,N+t-1)}, \hat{\psi}_{(\downarrow,N+t-1)}\}
    \end{align*}
    
    \State Initialize iteration counter $k \leftarrow 0$, error $\mathit{error}_1 \leftarrow \infty$, $\mathit{error}_2 \leftarrow \infty$, learning rate $\gamma \leftarrow \gamma_0$
    \While{$(\mathit{error}_1 > \epsilon_1 \ \mathbf{or} \ \mathit{error}_2 > \epsilon_2)$ \textbf{and} $k < M$}
      \State Orthogonalize and normalize $\psi_{(\uparrow, N+t)}^k$ and $\psi_{(\downarrow, N+t)}^k$ with respect to all orbitals in $S$
      \State Construct $H_{\mathit{CCSD}}$ for ($2N-1$)-electron system with $2(N+t)$ spin orbitals ($\psi_{(\uparrow, N+t)}, \psi_{(\downarrow, N+t)}$ and $S$)
      \State Compute gradients:
      \begin{equation*}
          \frac{\partial E_{\mathit{CCSD}}^k}{\partial \psi_{(\uparrow, N+t)}^k} \text{ and }
          \frac{\partial E_{\mathit{CCSD}}^k}{\partial \psi_{(\downarrow, N+t)}^k}
      \end{equation*}
      \State Update parameters:
      \begin{align*}
          \psi_{(\uparrow, N+t)}^{k+1} &\leftarrow \psi_{(\uparrow, N+t)}^k - \gamma\,\frac{\partial E_{\mathit{CCSD}}^k}{\partial \psi_{(\uparrow, N+t)}^k}\\
          \psi_{(\downarrow, N+t)}^{k+1} &\leftarrow \psi_{(\downarrow, N+t)}^k - \gamma\,\frac{\partial E_{\mathit{CCSD}}^k}{\partial \psi_{(\downarrow, N+t)}^k}
      \end{align*}
      \State Compute new energy error: $\mathit{error}_1 \leftarrow E_{\mathit{CCSD}}^{k+1} - E_{\mathit{CCSD}}^k$
      \State Compute new wavefunction error: $\mathit{error}_2 \leftarrow \max\big(\|\psi_{(\uparrow, N+t)}^{k+1} - \psi_{(\uparrow, N+t)}^k\|, \|\psi_{(\downarrow, N+t)}^{k+1} - \psi_{(\downarrow, N+t)}^k\|\big)$
      \If{$E_{\mathit{CCSD}}^{k+1} > E_{\mathit{CCSD}}^k$}
        \State Reduce learning rate: $\gamma \leftarrow \gamma/2$; \State Reset descent counter: $\mathit{consecutive\_descent} \leftarrow 0$
        \State \textbf{continue}
      \Else
        \State $\mathit{consecutive\_descent} \leftarrow \mathit{consecutive\_descent} + 1$
        \If{$\mathit{consecutive\_descent} > 5$}
          \State Slightly increase learning rate: $\gamma \leftarrow \gamma \times 1.1$
        \EndIf
      \EndIf
      \State Increment iteration counter: $k \leftarrow k + 1$
    \EndWhile
    \State Orthogonalize and normalize $\phi_{(\uparrow, N+t)}$ and $\phi_{(\downarrow, N+t)}$ with respect to all orbitals in $S$
    \State \textbf{return} optimized orbitals $\hat{\psi}_{(\uparrow, N+t)}$ and $\hat{\psi}_{(\downarrow, N+t)}$
  \end{algorithmic}
\end{algorithm}

\end{document}